Chapter 23

# Creation of Material Functions by Nanostructuring

**Marek Mezera[1], Camilo Florian[2], Gert-willem Römer[3], Jörg Krüger[1], Jörn Bonse[1,*]**

[1] Bundesanstalt für Materialforschung und -prüfung (BAM), Unter den Eichen 87, 12205 Berlin, Germany

[2] Princeton Institute for the Research and Technology of Materials (PRISM), Princeton University, Princeton, USA

[3] Chair of Laser Processing, Department of Mechanics of Solids, Surfaces & Systems (MS$^3$), University of Twente, 7500 AE, Enschede, The Netherlands

e-mails: camilofb1@gmail.com ; g.r.b.e.romer@utwente.nl ; joerg.krueger@bam.de ; joern.bonse@bam.de

*corresponding author

## Abstract

Surface nanostructures provide the possibility to create and tailor surface functionalities mainly via controlling their topography along with other chemical and physical material properties. One of the most appealing technologies for surface functionalization via micro- and nanostructuring is based on laser processing. This can be done either via direct contour-shaping of the irradiated material using a tightly focused laser beam, or in a self-ordered way that allows employing larger laser beam diameters along with areal scanning to create a variety of laser-induced periodic surface structures (LIPSS). For the latter approach, particularly ultrashort pulsed lasers have recently pushed the borders across long-lasting limitations regarding the minimum achievable feature sizes and additionally boosted up the production times. This chapter reviews the plethora of recently investigated applications of LIPSS - for example via imposing diffractive or plasmonic structural colors, the management of liquids and surface wetting properties, biomedical and bio-inspired functionalities, beneficial effects in tribology for reducing friction and wear, the manipulation of optical scattering and absorption in photovoltaics, or the manipulation of magnetic or superconducting surface properties in other energy applications. The footprint of the LIPSS-based technology is explored in detail regarding the current state of industrialization, including an analysis of the market and associated LIPSS production costs.

# 1. Introduction:
# Laser-induced Periodic Surface Structures – LIPSS

### 1.1 Zoology of LIPSS
### 1.2 Physical and Chemical Properties of LIPSS

# 2. Applications of LIPSS

### 2.1 Optical Applications
### 2.2 Liquid Management and Surface Wetting
### 2.3 Biological Applications
### 2.4 Tribological Applications
### 2.5 Energy Applications
### 2.6 Other Technical Applications

# 3. Industrialization

### 3.1 Market Analysis
### 3.2 Production Costs
### 3.3 Patent Situation

# 4. Outlook

# 1. Introduction: Laser-induced Periodic Surface Structures – LIPSS

*Laser-induced periodic surface structures* (LIPSS) were reported for the first time more than fifty years ago [Birnbaum, 1965]. Already during their early times, it became clear that LIPSS are a universal phenomenon that occurs when intense and coherent laser radiation is interacting with the surface of solids [van Driel, 1982]. Depending on the polarization state of the laser radiation (linear, circular, radial, vector states carrying an angular momentum, etc.), the formed LIPSS can exhibit very different surface morphologies and geometrical characteristics, e.g., spatial periods, modulation depths, aspect ratios, orientations, etc. (for a detailed classification of LIPSS see the following Section 1.1).

During the last decades, the topic of LIPSS has evolved into a scientific evergreen [Bonse, 2017]. While during the 1980s the fundamental theories on LIPSS were successfully developed [Sipe, 1983], the LIPSS phenomenon itself was rather considered as curiosity that finally did not entered the stage of practical applications. Later, starting around the turn of the millennium – with the broader availability of ultrashort pulsed laser systems in research laboratories – the topic has experienced again a vivid revival. From that time on, lots of scientific research has again focused on the fundamental formation mechanisms of LIPSS, particularly when created with femtosecond laser pulses. In that case, the stages of energy deposition to the solid driving the material into a strongly excited state, and the stage of subsequent matter re-organization are temporally well separated. For detailed review articles on the origin of LIPSS the reader is referred to [Siegman, 1986 / Bonse, 2012 / Vorobyev, 2013 / Buividas, 2014 / Abere, 2016 / Bonse, 2017 / Bonse, 2020a / Bonse, 2021] and to Chapter 1 (Derrien et al.), Chapter 2 (Ivanov et al.) and Chapter 5 (Rudenko et al.) of this book. Moreover, during the last five to ten years, a systematic screening of potential applications of LIPSS was performed [Vorobyev, 2013 / Liu, 2019 / Florian, 2020 / Gräf, 2020 / Bonse, 2020b / Stratakis, 2020 / Bonse, 2021], pointing towards many different scientific and practical aspects of LIPSS.

The following Chapter aims to provide a comprehensive review on the variety of different surface functionalizations that can be created through LIPSS. These specific surface functions finally can be transferred into various applications in optics, fluidics, medicine, tribology, energy saving, etc.

## 1.1 Zoology of LIPSS

In this Section the variety of different kinds of self-organized laser-induced surface structures is presented, providing a systematic and strict classification of LIPSS along with current terminologies that can be found in the literature. We focus particularly on *low spatial frequency LIPSS* (LSFL), *high spatial frequency LIPSS* (HSFL), *triangular nanopillars* (TNP) and explain how these sub-wavelength structures can be distinguished from supra-wavelength *Grooves* and more irregular *Spikes* structures.

Depending on the laser processing strategy (“spot processing” vs. “scanning processing” of lines or larger areas) and the specific laser irradiation parameters (fluence, number of pulses, spatial beam profile, etc.), structures of a single type of LIPSS, spatially separated regions of distinct LIPSS, or even multiple mixed types of LIPSS or hierarchical surface structures may be simultaneously observed [Bonse, 2021].

In a strict classification, LIPSS represent a 1D- or sometimes 2D-quasi-periodic grating structure and are usually classified with respect to their spatial period $\Lambda$ (compared to the irradiation wavelength $\lambda$), their relation to the laser beam polarization (linear, circular, etc.), and their depth-to-period aspect ratio $A$.

## Low Spatial Frequency LIPSS (LSFL)

The most prominent type of LIPSS that was historically observed first in 1965 by Birnbaum [Birnbaum, 1965] are LSFL. For radiation that is incident normal to the surface of a solid (angle of incidence $\theta = 0°$), LSFL typically have periods close to or somewhat smaller than the laser irradiation wavelength $\lambda$ used to create them ($\lambda/2 \leq \Lambda_{LSFL} \leq \lambda$) and an aspect ratio $A$ typically < 1. These structures usually form in the ablative regime at fluence levels up to several times the ablation threshold (material specific), although also melting-induced or oxidative LSFL were reported [Bonse, 2017]. For strongly absorbing materials such as metals or semiconductors, LSFL are typically oriented perpendicular to the linear laser beam polarization (type LSFL-I), while for weakly absorbing wide band gap dielectrics, they are typically parallel to it (type LSFL-II). Note that the numbering of the different types of LIPSS with Roman letters (-I/-II) follow the historic order of observation.

The dissimilarity of both types of LSFL arises from the different coherent optical scattering mechanisms involved in their formation. LSFL-I originate from the excitation of so-called *Surface Electromagnetic Waves* (SEWs) [Akhmanov, 1985 / Bonch-Bruevich, 1992], such as *Surface Plasmon Polaritons* (SPPs) and the optical interference of their electromagnetic fields with the incident laser beam itself, resulting in a spatially modulated pattern of the locally deposited optical energy that finally translates into a periodically corrugated surface topography [Bonse, 2009 / Bonse, 2020a]. For p-polarized laser radiation, the surface grating relief then exhibits spatial periods close to the wavelength ($\Lambda_{LSFL\text{-}I,p} \sim \lambda/[1\pm\sin(\theta)]$). In contrast, for dielectrics, the LSFL-II are seeded some hundreds of nanometers below the surface and originate from the interference of the incident (propagating) laser radiation with light coherently scattered at the surface roughness (via radiative fields that are propagating towards the far-field region) [Rudenko, 2017]. Via multiple laser pulse irradiation and incubation effects in the material (inter-pulse feedback), the LSFL-II signature emerges at the surface once the covering near-surface material is removed through ablation ($\Lambda_{LSFL\text{-}II} \sim \lambda/\{n[1\pm\sin(\theta)]\}$, with $n$ being the (real valued) refractive index of the dielectric [Siegman, 1986]).

Figure 1 exemplifies LSFL-I that were formed at the surface of a niobium metal sheet upon scanning fs-laser pulse irradiation in air environment [790 nm, 30 fs, 1 kHz, [Cubero, 2020a]]. Fig. 1(a) shows a scanning transmission electron micrograph (STEM) of a cross-sectional cut through the surface of the processed LSFL. It can be observed that the LSFL-I are almost sinusoidal with a period of ~570 nm and a modulation depth of ~300 nm. Note the fine boundary (marked by five white arrows) that separates the laser-affected region from the non-affected poly-crystalline niobium bulk material underneath. Supposedly, the niobium surface was melted up to this boundary during the fs-laser scan processing, resulting in a re-solidified superficial layer of 40 to 300 nm here [Cubero, 2020a]. Figure 1(b) provides a corresponding top-view scanning electron micrograph of the LSFL-I.

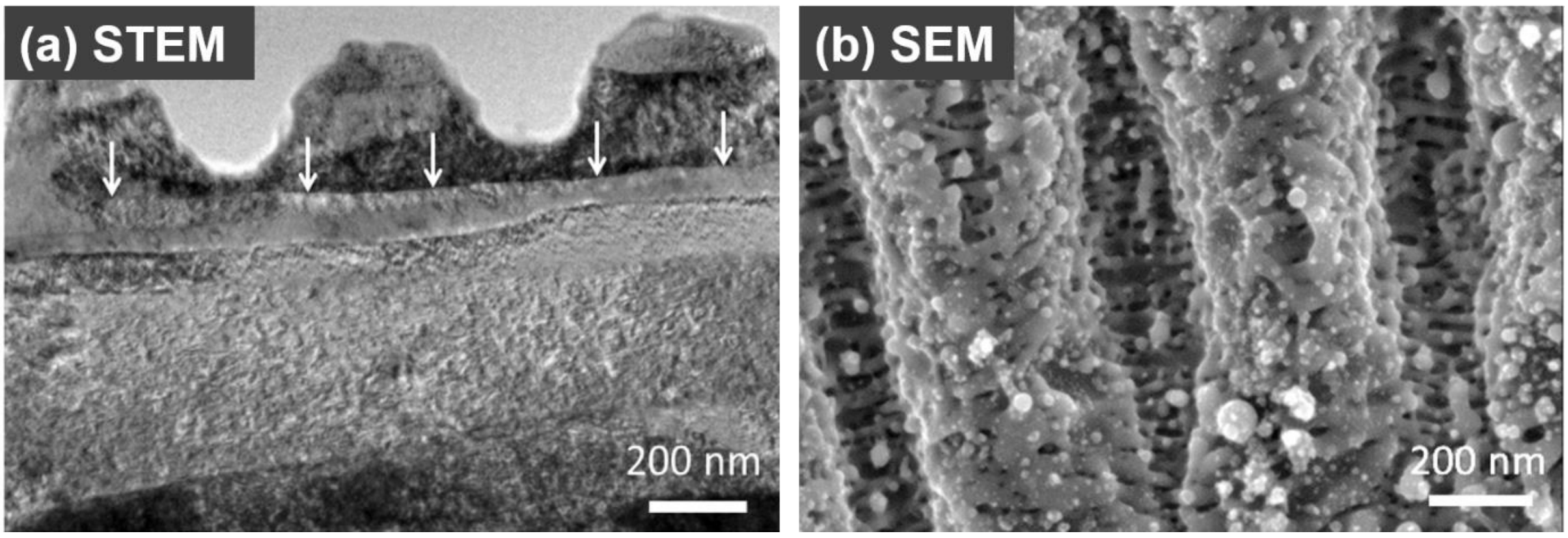

Fig. 1: (a) STEM micrograph of the cross-sectional surface profile of LSFL-I on poly-crystalline niobium [790 nm, 30 fs, 1 Hz]. (b) Top-view SEM micrograph of the corresponding surface structures. The white arrows in (a) mark at the boundary between laser-affected zone and non-affected bulk material. (Reprinted (adapted) from [Cubero, 2020a], Cubero et al., Surface superconductivity changes of niobium sheets by femtosecond laser-induced periodic nanostructures, Nanomaterials (Basel, Switzerland) **10**:2025, Copyright 2020 under Creative Commons BY 4.0 license. Retrieved from https://doi.org/10.3390/nano10122525).

## High Spatial Frequency LIPSS (HSFL)

HSFL typically have periods smaller than half of the irradiation wavelength ($\Lambda_{HSFL} < \lambda/2$) and are oriented either parallel or perpendicular to the laser beam polarization (materials dependent). They are formed at fluence levels very close to the damage threshold of the irradiated material and predominantly for ultrashort pulse durations in the fs- to ps-range. Two types are usually distinguished.

HSFL-I are mainly observed on transparent materials (dielectrics and semiconductors). They consist of very narrow periodic grooves having individual widths of a few tens of nanometers only, while their depth can reach up to several hundreds of nanometers. Given their large aspect ratio $A > 1$, some authors call these nanostructures *"deep-subwavelength ripples/structures"*. For (semi)transparent materials and normal incident laser radiation HSFL-I usually exhibit spatial periods of $\Lambda_{HSFL\text{-}I} \sim \lambda/(2n)$, with *n* being the refractive index of the material. HSFL-I originate from the near-(sub)-surface interference of the incident (propagating) laser radiation with light coherently near-field scattered at nanoscale surface corrugations (via non-radiative fields confined to the near-field region of the scattering centers) [Rudenko, 2017].

Figure 2 demonstrates the large depth-to-period aspect ratio *A* of the HSFL-I (b,c) that were formed in the periphery of an LSFL-I covered ablation crater (a) at the surface of a silicon carbide (SiC) crystal after spot processing by fs-laser pulses in air [800 nm, 150 fs, 50 Hz, [Tomita, 2009]].

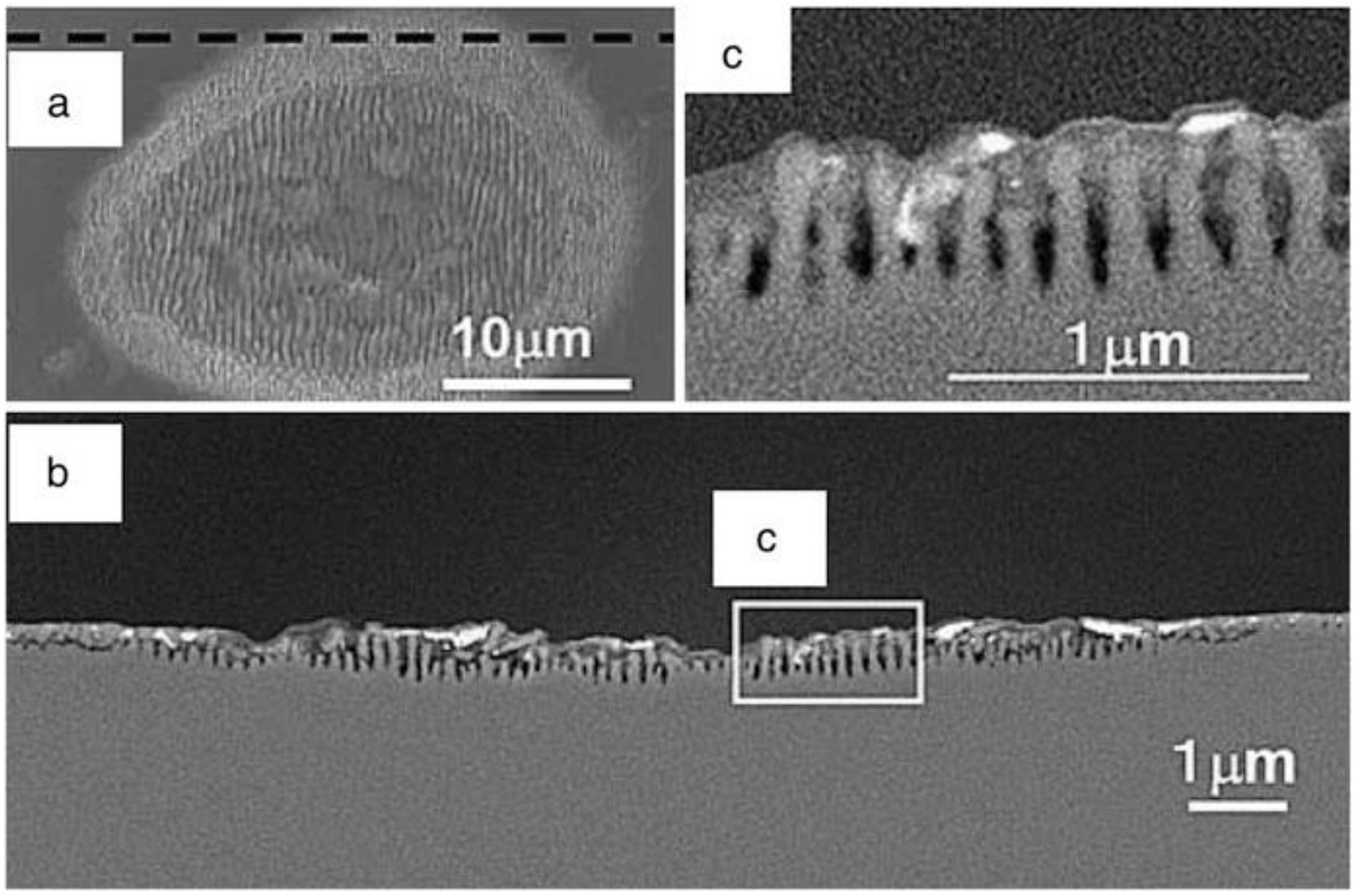

Fig. 2: (a) Top-view SEM image LSFL-I (center) and HSFL-I (periphery) on SiC. The dashed black line marks the position where the cross-sectional profiles (b,c) were acquired [800 nm, 150 fs, 50 Hz]. (b) Cross-sectional SEM micrograph of the whole irradiated spot. (c) Detailed magnification of selected area (c) marked in (b). Taken from [Tomita, 2009]. (Reprinted by permission from Springer-Verlag: Applied Physics A **97**:271–276 (Cross-sectional morphological profiles of ripples on Si, SiC, and HOPG, Tomita, T. et al.), Copyright (2009)).

In contrast, HSFL-II have depths of only a few tens of nanometers and spatial periods of the order of ~100 nm. Hence, the resulting aspect ratio is far smaller than one ($A \ll 1$). These shallow structures are often observed on metal surfaces, and different formation mechanisms were suggested, including surface oxidation or twinning effects.

From the three observations that both types of HSFL mainly form upon irradiation with ultrashort laser pulses, are strictly related to the direction of the linear laser beam polarization, and have a lower period limit of a few tens of nanometers (caused by thermal diffusion effects washing out too small spatial modulations during the electron-phonon relaxation stage), it was concluded that ultrafast energy deposition dominates their formation [Bonse, 2017]. However, for the HSFL-II, recent numerical 3D *Finite-Difference Time-Domain* (FDTD) calculations [Skolski, 2012] combined with a numerical solution of the *Navier-Stokes equations*, point towards a complex interplay of coherent electromagnetic scattering and subsequent thermoelastic and hydrodynamic effects, all triggered by the strong ultrafast laser excitation of the solid [Rudenko, 2020] (see also **Chapter 5** (Rudenko et al.)).

## Triangular Nanopillars (TNP)

Starting in 2018, another characteristic surface morphology was reported by several groups for the irradiation of different metals by ultrashort laser pulses in a scan-processing geometry under specific laser processing conditions [Romano, 2018 / Liu, 2018a / Fraggelakis, 2018 / van der Poel, 2019, Mezera, 2020]. These surface structures consist of nanometric triangular-shaped nanopillars as building blocks for hexagonally arranged meta-units that finally form a large area 2D-grating at the laser-processed surface. Therefore, this type of LIPSS is referred to either as *"triangular nanopillars"* (TNPs) or as "*hexagonally arranged nano-triangles*" in the pertinent literature. While in Romano et al. [Romano, 2018] the TNPs were realized on stainless steel with multiple circularly polarized fs-laser pulses at high repetition rate (1032 nm, 310 fs, 250 kHz), Liu et al. [Liu, 2018a] reported their formation on tungsten surfaces upon low repetition rate (800 nm, 50 fs, 1 kHz) double-fs-pulse irradiation using cross-polarized linear polarization and an inter-pulse delay of ~1 ps and with cylinder lens focusing. Fraggelakis et al. [Fraggelakis, 2019] used high repetition rate (1030 nm, 350 fs, 100 kHz) double-fs-pulse with circular counter rotating polarizations and inter-pulse delay up to a few nanoseconds and with spherical lens focusing for their processing. Later, van der Poel et al. [van der Poel, 2019] generated these TNPs on a cobalt–chrome–molybdenum–alloy (CoCrMo) with multiple circularly polarized ps-laser pulses (1030 nm, 6.7 ps, 400 kHz). An example from that work is provided in Fig. 3, where two top-view SEM micrographs (a,b) are shown that were taken at different magnifications. Figure 3(c) is a 2D-Fast Fourier Transform (2D-FFT) of the micrograph displayed in (a), representing a 2D-histogram (map) of the spatial frequencies contained in the transformed image. The characteristic arrangement of the pronounced peaks in Fig. 3(c) reflects the hexagonal symmetry of the TNP-grating structure, while their positions indicate very similar spatial periods of $\Lambda_{TNP}$ ~860 nm found in several directions here.

The physical origin of the TNPs is currently vividly discussed in the scientific community. Since their spatial periods are very close to that of the LSFL-I when formed at the same material ($\Lambda_{TNP} \approx \Lambda_{LSFL\text{-}I}$), it is reasonable that a SPP-based formation mechanism should be invoked for their explanation [Liu, 2018a / Makin, 2020 / Zhang, 2020]. Additionally, a nonlinear hydrodynamic convection flow mechanism was proposed to be relevant [Fraggelakis, 2019]. (see also **Chapter 5** (Rudenko et al.)).

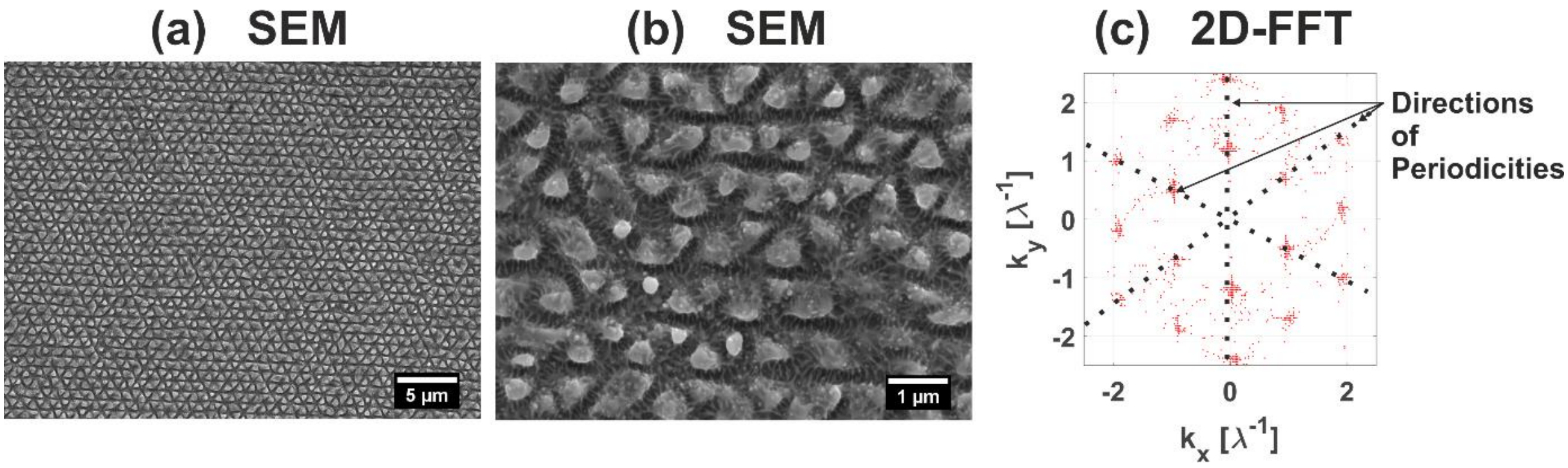

Fig. 3: (a,b) Top-view SEM micrographs of TNPs processed on CoCrMo-alloy upon scan-processing with circularly polarized ps-laser pulses [1030 nm, 6.7 ps, 400 kHz]. (c) 2D-FFT of the micrograph shown in (a) indicating a hexagonal symmetry of the TNP arrangement along with spatial periods of $\Lambda_{TNP}$ ~ 860 nm. (Adapted and reprinted from [van der Poel, 2019], van der Poel et al., Fabricating Laser-Induced Periodic Surface Structures on Medical Grade Cobalt–Chrome–Molybdenum: Tribological, Wetting and Leaching Properties, Lubricants (Basel, Switzerland) **7**:70, Copyright 2019 under Creative Commons BY 4.0 license. Retrieved from https://doi.org/10.3390/lubricants7080070).

## Classification of Laser-Generated Surface Structures

Figure 4 orders the plethora of different laser-generated surface structures in a single scheme. The branch of the LIPSS was discussed already in detail in the previous sub-sections. Apart from the LIPSS, also supra-wavelength *Grooves* ($\Lambda_{Grooves} > \lambda$) were observed, typically featuring an orientation parallel to the laser beam polarization and periods in the few micrometers range. Even larger structural sizes are observed for the irregular *Spikes* morphology (sometimes referred to as *Micro-Cones*) which exhibit a very weak correlation to the polarization only and "periods" up to some tens of micrometers ($\Lambda_{Spikes} >> \lambda$).

The general requirements on the laser fluence and the number of laser pulses for the generation of the different surface nano- and micro-structures are indicated by the arrow in the bottom part of Fig. 4. For a more detailed discussion of the specific formation mechanisms of Grooves and Spikes the reader is referred to the specialized literature [Stratakis, 2020 / Bonse, 2020a / Mezera, 2020 / Nivas, 2021].

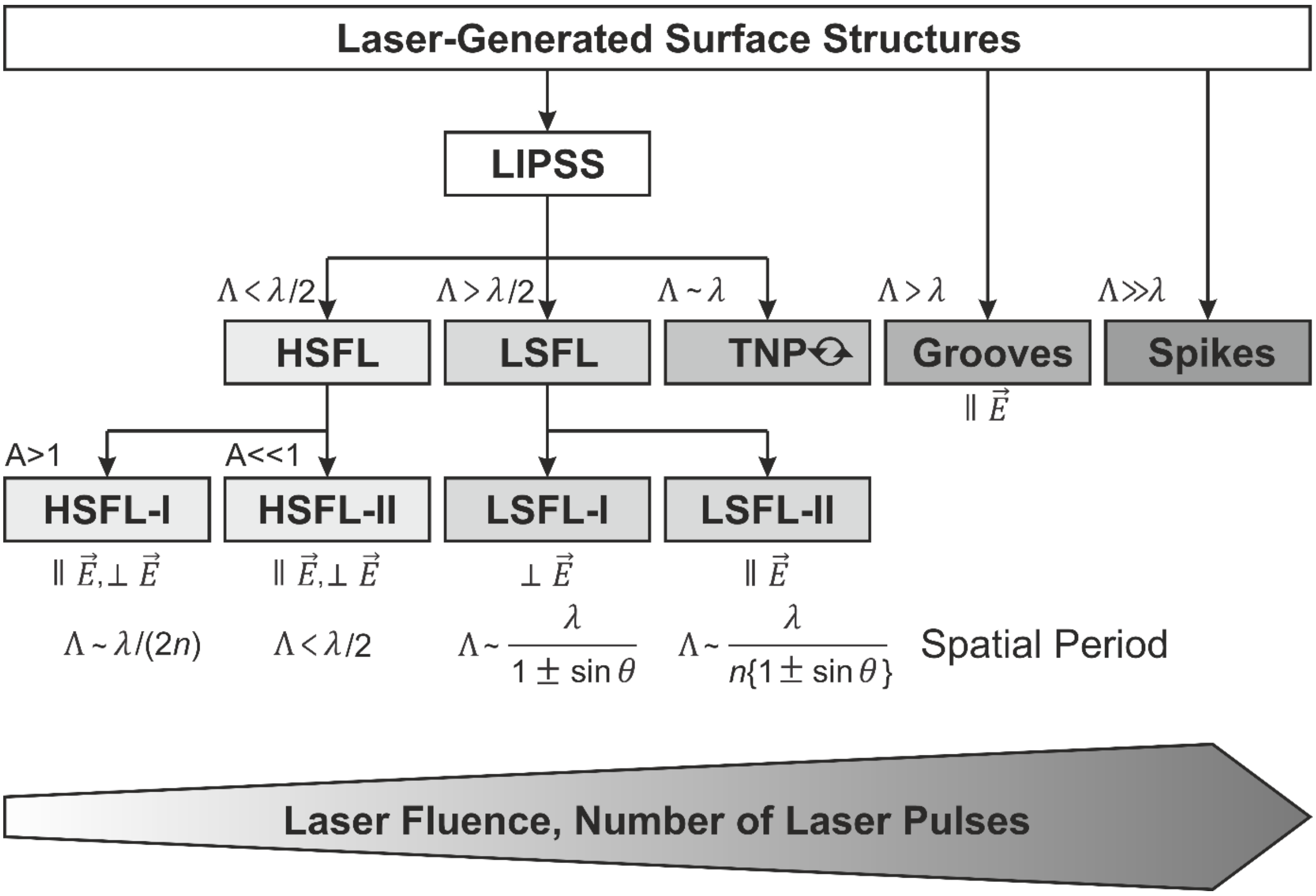

Fig. 4: Classification of laser-generated surface structures. Adapted from [Mezera, 2020].

## 1.2 Physical and Chemical Properties of LIPSS

In this Section physical and chemical surface properties are discussed that may be altered and controlled through LIPSS for creating specific surface functions. Figure 5 summarizes the different properties that are discussed in more detail in the following two sub-sections *Physical Properties* and *Chemical Properties*.

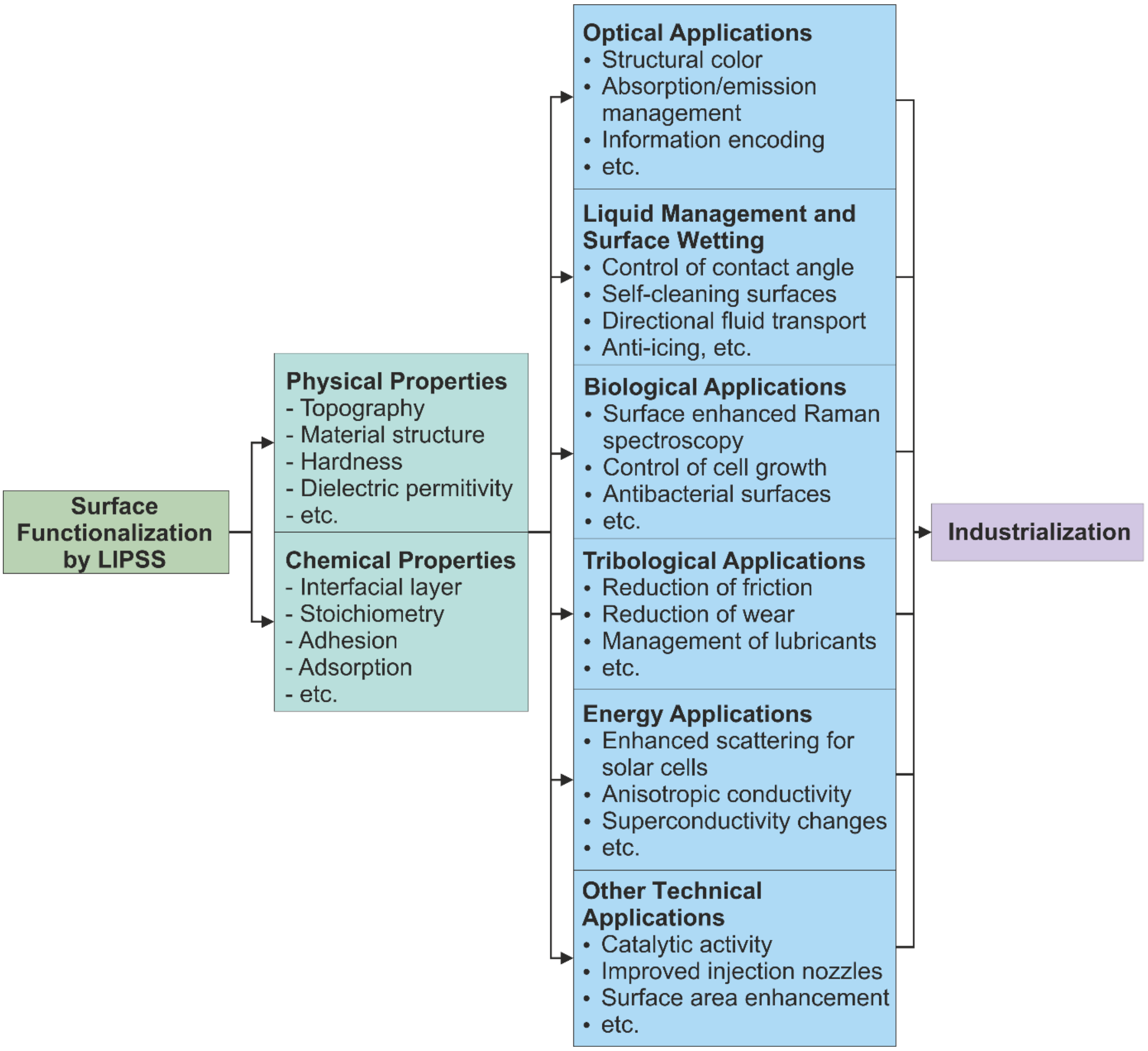

Fig. 5: Scheme of properties affected by LIPSS and resulting applications for surface functionalization, providing the frame for this book chapter. Inspired by [Florian, 2020a].

## Physical Properties

LIPSS can create various specific physical surface properties. The most obvious one is the characteristic grating-like *surface topography* that is describing the interface between the sample material and the ambient environment. Since the formation of LIPSS includes a complex chain of different stages, starting with the optical material excitation via absorption by the electrons of the solid, followed by the transfer of the energy from the electrons to the lattice of the solid, and a cascade of subsequent relaxation processes, such as melting, ablation, solidification in amorphous or crystalline states, etc. [Bäuerle, 2011 / Bonse, 2020a], it becomes obvious that also structural material properties can be altered through the laser-processing of LIPSS. As consequence, e.g., the materials hardness [Bonse, 2018], the dielectric permittivity [Hwang, 2010], the electric conductivity [Lopez-Santos, 2021], surface superconductivity [Cubero, 2020a], etc. can be locally modified through the presence of LIPSS, featuring specific applications in tribology, photonics, or electronics [Florian, 2020a / Gräf, 2020 / Bonse, 2021].

### Chemical Properties

Apart from changes of physical properties, the formation of LIPSS is usually accompanied by chemical surface alterations, particularly if the laser-processing is performed in air environment – a requirement that is strongly demanded for large area or industrial laser processing. Most prominent, typically surface oxidation occurs since high transient temperatures are reached in the reactive atmosphere air upon the processing of LIPSS [Kirner, 2017a / Kirner, 2018 / Florian, 2020b / Florian, 2020c]. Hence, in an interfacial surface layer, the chemical composition of the irradiated material may change.

Apart from intrinsic stoichiometric material alterations, other topography- and roughness-related effects can be rendered possible through LIPSS: the enlarged surface area of the LIPSS allows molecules, such as hydrocarbons, to adsorb from the ambient environment [Kietzig, 2009 / Yasumaru, 2017 / Gregorčič, 2021]. Through such "ageing" effects, even a transition from an initially hydrophilic surface into a superhydrophobic state can manifest – without any significant change of the surface topography. For many biomedical or healthcare applications, such topographic and surface wetting effects are particularly relevant for the controlled adhesion (repellent or adducent) of specific biological cells or even for the prevention of the formation of bacterial biofilms [Larrañaga-Altuna, 2021 / Zheng, 2021]. On the other hand, the LIPSS-enhanced total surface area can help to increase the efficiency of catalytic chemical reactions [Lange, 2017].

From all these effects, different fields of surface functionalization can be identified that are discussed in the following sub-Chapter 2 *Application of LIPSS*, including *Optical Applications* (Section 2.1), *Liquid Management and Surface Wetting* (Section 2.2), *Biological Applications* (Section 2.3), *Tribological Applications* (Section 2.4), *Energy Applications* (Section 2.5), and *Other Applications* (Section 2.6). Finally, sub-Chapter 3 analyses the current state of *Industrialization* of LIPSS, providing a *Market Analysis* (Section 3.1), some quantitative estimates of the *Production Costs* (Section 3.2), and a compilation of the actual LIPSS-related *Patent Situation* (Section 3.3).

# 2. Applications of LIPSS

As outlined in the previous sub-Chapter, LIPSS exhibit some very specific intrinsic properties that can be used to create and tailor very different surface functionalities. Most of them finally fall back either to the topographic characteristics of LIPSS, or to the specific surface chemistry involved during their formation (or even later). These surface functionalities enabled various applications that were studied already in detail in the context of LIPSS.

## 2.1 Optical Applications

In this Section optical applications of laser-induced nanostructures on surfaces and sub-surface layers are addressed. It is discussed that laser-induced periodic surface structures (LIPSS, ripples), nanostructure-textured microgrooves, and nanospikes can lead to diffractive structural coloring effects on metals (via ripples), modify the optical absorption of metal or semiconductor surfaces towards "black" materials (via microgrooves) or can decrease the reflectance on dielectrics leading to an anti-reflection effect (via nanospikes). Apart from optical emission and absorption management, safety tags can be realized via near surface 3D-self-organization of nanocomposite films.

## Structural Color

LIPSS with periods of the order of visible wavelengths can efficiently act as optical diffraction gratings and, thus, generate structural colors for optical applications. While some applications require the processing of individual spots or tailored lines, most of these utilizations rely on the homogeneous processing of LIPSS on large surface areas. Usually, this is implemented through laser scanners realizing a relative motion between the focused laser beam and the workpiece.

Figure 6 exemplifies the processing of spots, lines, and areas on a polished steel surface, see the scheme in Fig. 6(a). For a fixed laser source (wavelength and pulse duration), the period of the LIPSS can be controlled by the laser fluence ($\phi_0$) and the effective number of laser pulses per beam spot diameter ($N_{\text{eff}}$). On the sample, individual spots and lines connecting the areas generated through a meandering line-wise motion were processed by a Ti:Sapphire fs-laser system. Depending on the LIPSS period, different structural colors appear as result of optical diffraction upon illumination of the sample with a white light source. This is demonstrated in Fig. 6(b), where color effects can be seen for the structured areas, lines and even spots.

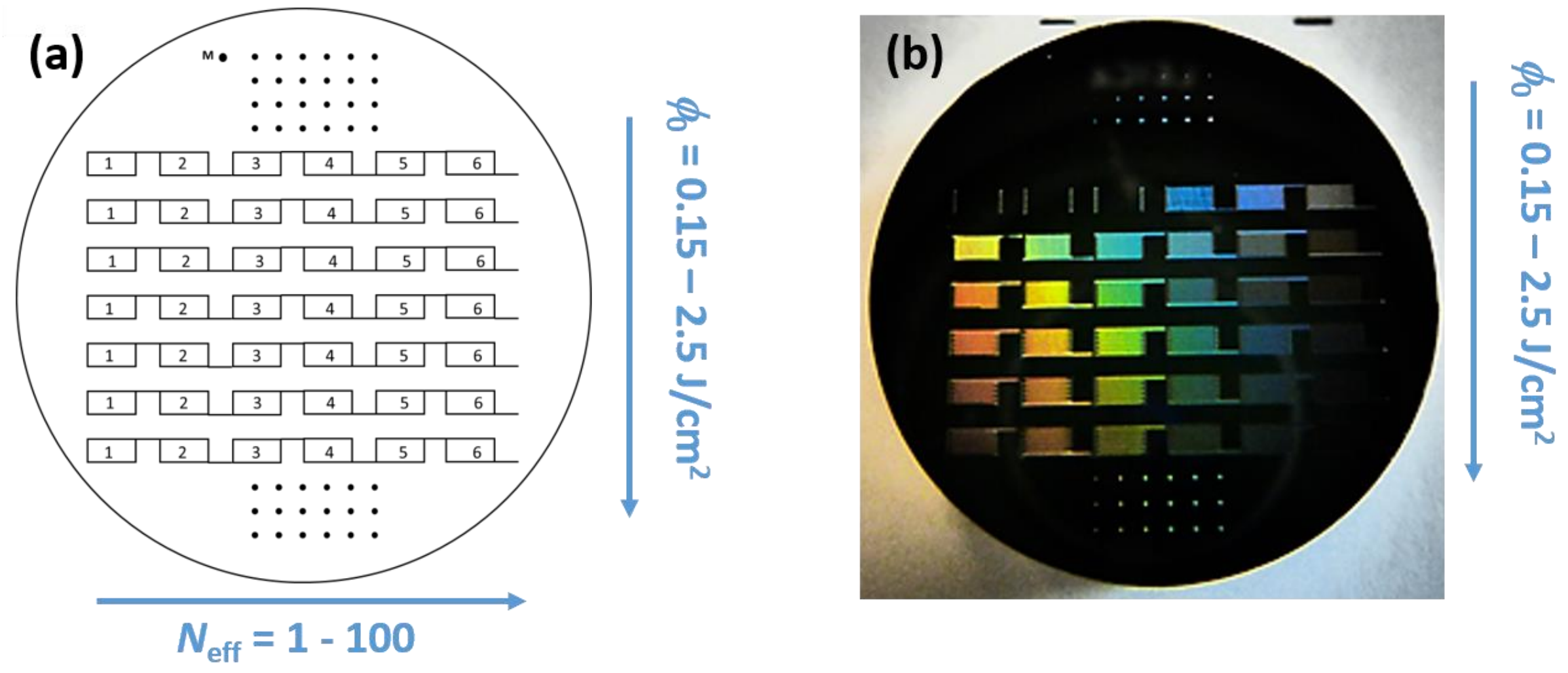

Fig. 6: (a) Scheme of laser processing. (b) Photograph of a polished 100Cr6 steel sample (24 mm diameter) after fs-laser processing [800 nm, 30 fs, 1 kHz] of spots, lines, and areas. The colors arise from optical diffraction of ambient light at laser generated LSFL. (© 2016 IEEE. Reprinted, with permission, from [Bonse, 2017], Bonse, J., et al., Laser-induced periodic surface structures—A scientific evergreen, IEEE J. Sel. Top. Quantum Electron, 2017, **23**:9000615).

A pioneering work demonstrating the application of LIPSS for color marking and decoration purposes was presented by Dusser et al. [Dusser, 2010] encoding a miniaturized, colored picture of the painter van Gogh on a steel surface, see Fig. 7. The color variation was implemented by a variation of the orientation of LSFL for a fixed angle of observation here.

For a more detailed discussion of structural color the reader is referred to the recent review article of Liu et al. [Liu, 2019] and the references included.

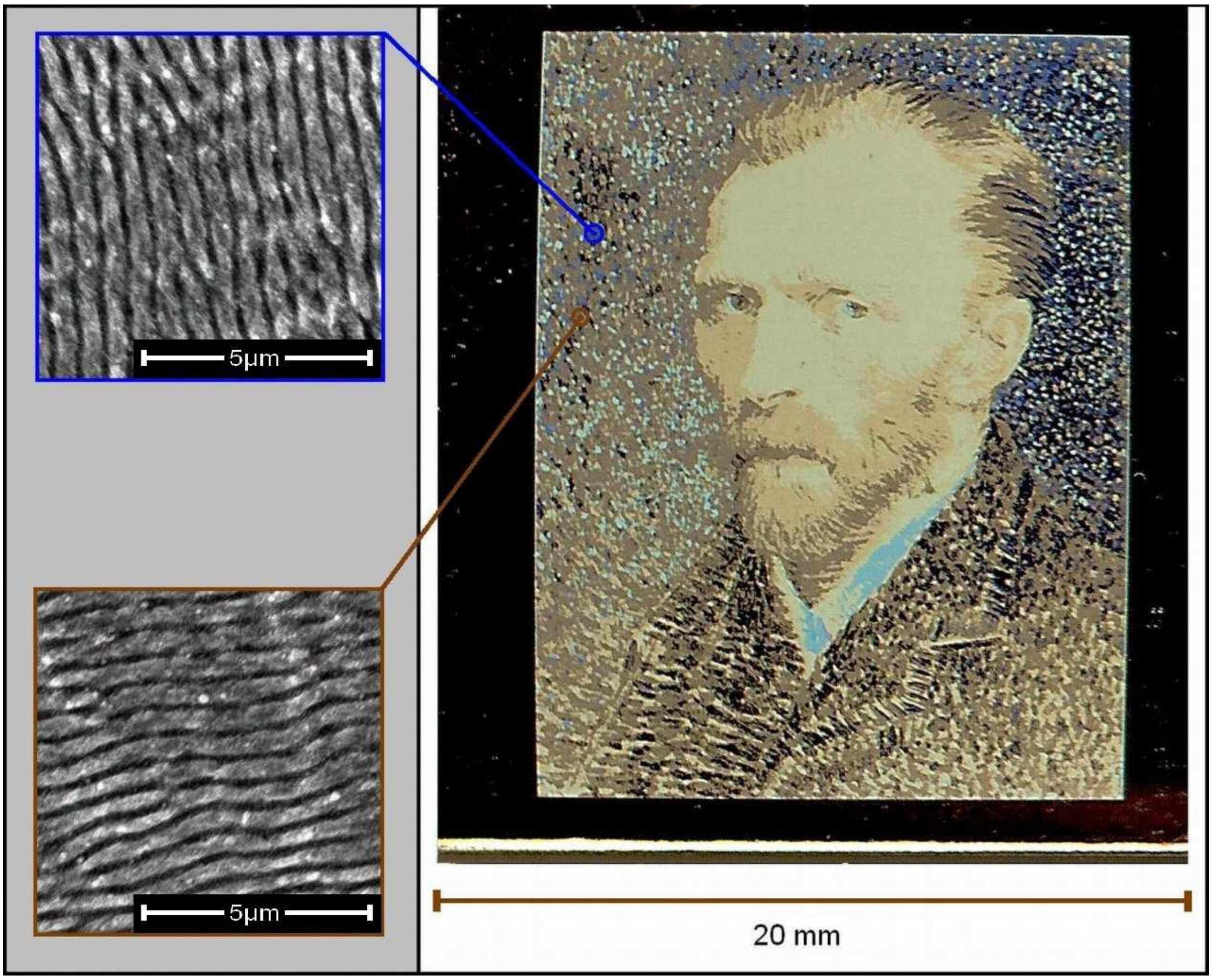

Fig. 7: Photograph (right) of a polished 316L stainless steel sample after fs-laser processing [800 nm, 150 fs, 5 kHz] of a miniaturized self-portrait of the painter Vincent van Gogh. The insets (left) are top-view scanning electron micrographs of LSFL with two different local orientations. (Reprinted with permission from [Dusser, 2010], © Optical Society of America).

## Absorption and Emission Management

Hierarchical micro-nanostructures (e.g. microgrooves and spikes) largely increase the effective surface of a sample. Due to locally non-normal incident rays and multiple reflections, these surface topographies can cause an increase in the absorption of light. This effect is visualized in Fig. 8 for nanostructure-textured microgrooves on platinum forming a "black metal" [Vorobyev, 2013]. The figure compares the spectral dependence of the reflectance of the polished Pt surface (full squares) with that of a fs-laser processed black Pt surface (full circles). The latter surface exhibits hemispherical reflectances of less than 5% in the ultraviolet to infrared spectral range (~250 nm to 2500 nm). The inset provides a picture of the corresponding laser-processed black Pt sample.

Regarding the emission management, Kirchhoff's law can be exploited. It states that the emittance of a surface equals its absorbance in thermal equilibrium. Hence, surfaces of increased absorption simultaneously feature an elevated emission. The effect can be used e.g. for filaments of incandescent lamps. This application has been demonstrated in a tungsten filament lamp increasing its optical emission efficiency by a factor of three in the visible range compared to an untreated filament [Vorobyev, 2009a].

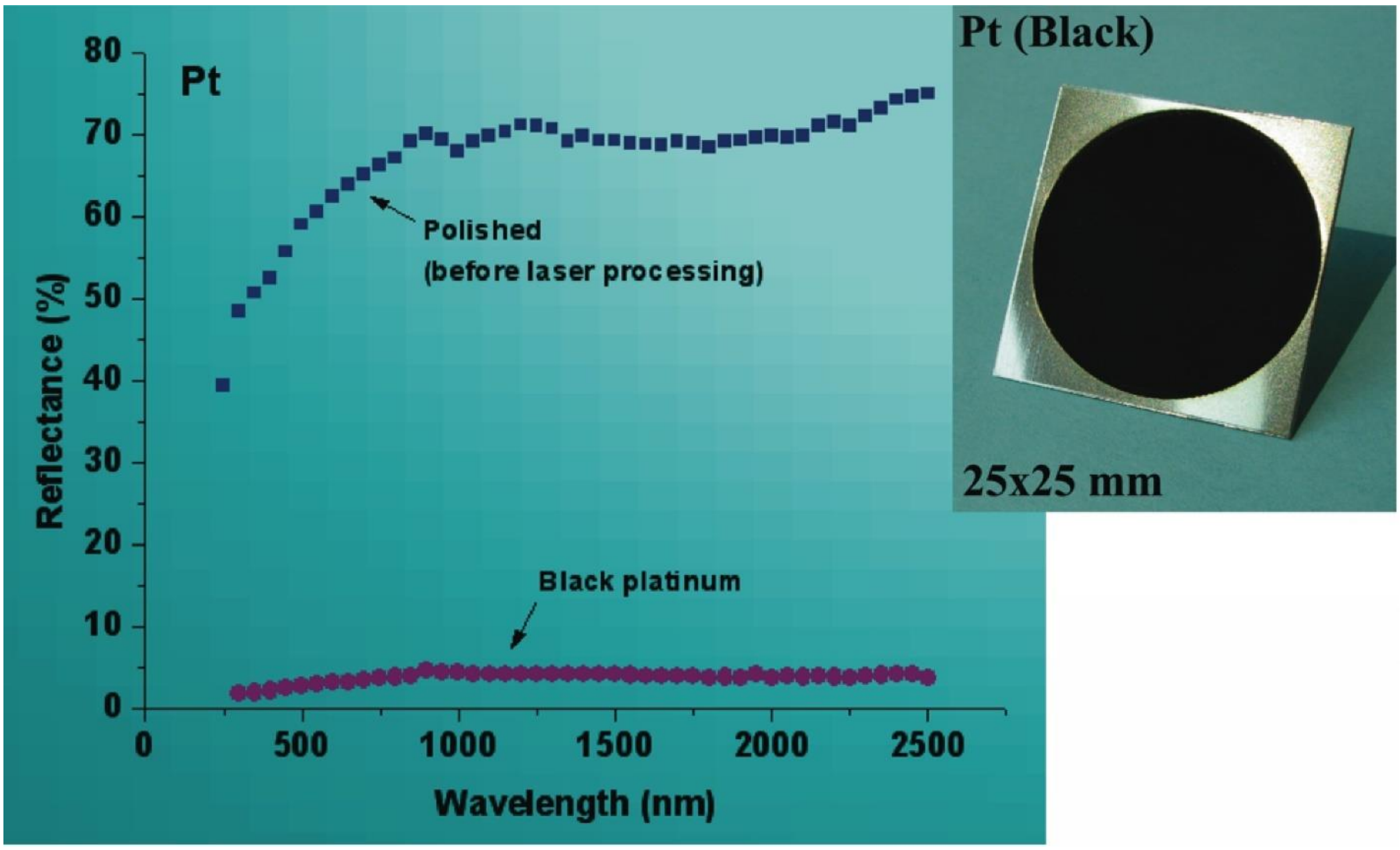

Fig. 8: Photograph of a fs-laser processed [800 nm, 65 fs, 1 kHz] "black platinum" sample covered with nanostructure-textured microgrooves (right). The corresponding total hemispherical reflectance spectrum is shown (left, violet full circles) along with a measurement of the polished surface as a reference (left, blue full squares). (Reprinted from [Vorobyev, 2013] by permission from Wiley-VCH-Verlag GmbH: Laser Photonics Rev. **7**:285–307 (Direct femtosecond laser surface nano/microstructuring and its applications, Vorobyev, A.Y. and Guo, C.), Copyright (2012)).

For transparent materials optical scattering effects can be used to create omnidirectional broadband antireflective surfaces [Stratakis, 2020]. Papadopoulos et al. [Papadopoulos, 2019] used circularly polarized fs-laser pulses to generate randomly arranged nanospikes for mimicking the wing of the butterfly *Greta Oto*. These laser-generated structures exhibit random height and widths distributions in the few hundreds of nanometer range, resulting in surface reflectance levels down to 1% in the visible spectrum.

## Information Encoding

Different technical approaches were employed to realize the encoding of information through self-organized nanostructures, e.g. for security tag applications. Apart from the obvious application of using a long-lasting structural coloring to avoid falsification of (Canadian) coins [Guay, 2017], more complex laser processing strategies were suggested by the groups around N. Destouches and J. Siegel [Liu, 2017 / Sharma, 2019a / Sharma, 2019b]. The technique applies silver salts embedded into (meso-)porous titania films to create layered and LIPSS-periodically arranged Ag-nanoparticle systems that can be used for multiplexed image encoding for brand identification, authentication of goods, and smart visual effects [Sharma, 2019a].

Figures 9(a,b,c) sketches the encoding of two distinct grating structures that can be processed in different depths in a mesoporous Ag:$TiO_2$-film on glass upon high-repetition rate fs-laser

pulse exposure (500 kHz) at different scanning speeds, i.e. “low speed” (LS) and “high-speed” (HS). The gratings arise from a plasmonically driven precipitation of Ag nanoparticles within the titania matrix, along with Ostwald ripening effects, resulting in a complex 3D-arrangement of differently sized Ag nanostructures (for details see [Liu, 2017]). These effects can be used for encoding several pictures in a multiplexed way, that can be optically read out under different rotational angles, see Figs. 9(d,e,f). Such an image encoding could be used for ID cards, etc.

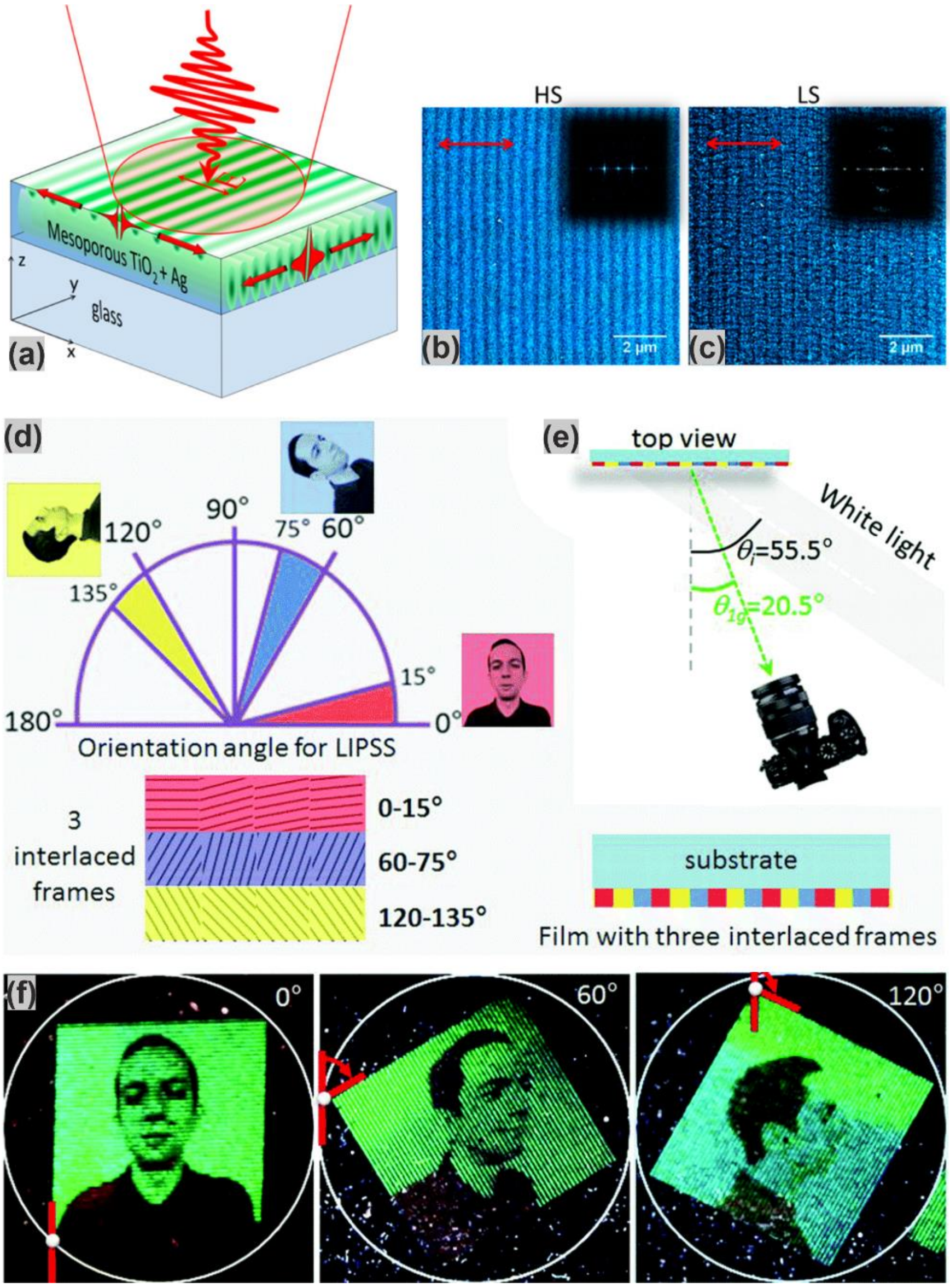

[Fig. 9: (a) Scheme of laser-induced 3D-self-organization in the mesoporous Ag:$TiO_2$ nanocomposite film. Linearly polarized laser pulses [515 nm, 370 fs, 500 kHz] excite simultaneously a surface wave and a guided wave, both propagating in perpendicular directions. The waves interfere with the incident pulse and cause an intensity modulation either close to the film surface parallel to the polarization, or within the film in the perpendicular to it. Both interference patterns feature different spatial periods. (b,c) Top-view micrographs of structures “HS” (b; processed with high speed of 100 mm/s) and “LS” (c; processed with a low speed of 10 mm/s) and their respective 2D-Fourier transforms (inset). The linear polarization direction is indicated by the red double arrows. (Images (a)-(c) reprinted with permission from [Liu, 2017] (Liu et al., Three-dimensional self-organization in nanocomposite layered systems by ultrafast laser pulses, ACS Nano **11**:5031–5040). Copyright (2017) American Chemical Society). (d,e,f) Inscription of multiple “hidden” images. (d) Angular ranges of LIPSS used to encode the 256 grey-levels of each image. The three raster images have been interlaced and printed as different fames onto a 1.18 $cm^2$ surface. The red, blue and yellow colors represent false colors used for interlacing. The three frames contain gratings whose orientation varies (red:0°–15°; blue: 60°–75°; yellow:120°–135°. (e) Scheme of the setup used for taking photographs of the sample, indicating the incidence angle $\theta_i$ of the white light collimated beam and the angle $\theta_{1g}$ of the first diffraction order in the plane of incidence for green light at a wavelength of 537 nm. (f) Photographs of the sample printed onto a polycarbonate substrate recorded for three different azimuthal angles upon rotating the sample in its plane. Three different images appear at 0°, 60° and 120°, respectively. (Images (d)-(f) republished with permission of Royal Society of Chemistry, from [Sharma, 2019a], Laser-driven plasmonic gratings for multiple image hiding, Sharma, N., et al., Mater. Horiz. **6**: 978–983, 2019; permission conveyed through Copyright Clearance Center, Inc.).

## 2.2 Liquid Management and Surface Wetting

In this Section wetting and liquid management applications of laser-induced micro- and nanostructures on surfaces are addressed. That is, besides surface chemistry, specific surface morphologies, including laser-induced periodic surface structures (LIPSS, ripples), can be exploited to tune and optimize the wetting properties of a surface. First the physics governing surface wetting are addressed briefly. Next examples are discussed showing that surface textures (including LIPSS) can be used to create surfaces which are self-cleaning, or allow to direct fluids, show reduced adhesion to liquid food, reduce drag, reduce the *Leidenfrost temperature*, and last but not least, show anti-icing properties.

### Physics Governing Surface Wetting

Wettability of a solid surface is the degree to which a surface can get wet by a liquid. The wetting properties of a liquid (droplet) on a solid surface, in a gaseous environment (a so-called solid–liquid–vapour system) are typically quantified by contact angle measurements, see Fig. 10(a). In this figure, the balance of capillary forces acting at the liquid–solid–vapour interface, results in the so-called *Young's contact angle* $\theta_Y$ [Young, 1805], which is related to the interfacial surface tension $\gamma_{vs}$ of the vapour-solid, $\gamma_{sl}$ of the solid-liquid and $\gamma_{lv}$ of the liquid-vapour, which are related as $\cos(\theta_Y) = (\gamma_{vs}-\gamma_{sl})/\gamma_{lv}$. The interfacial surface tensions are chemical material parameters [Samanta, 2020].

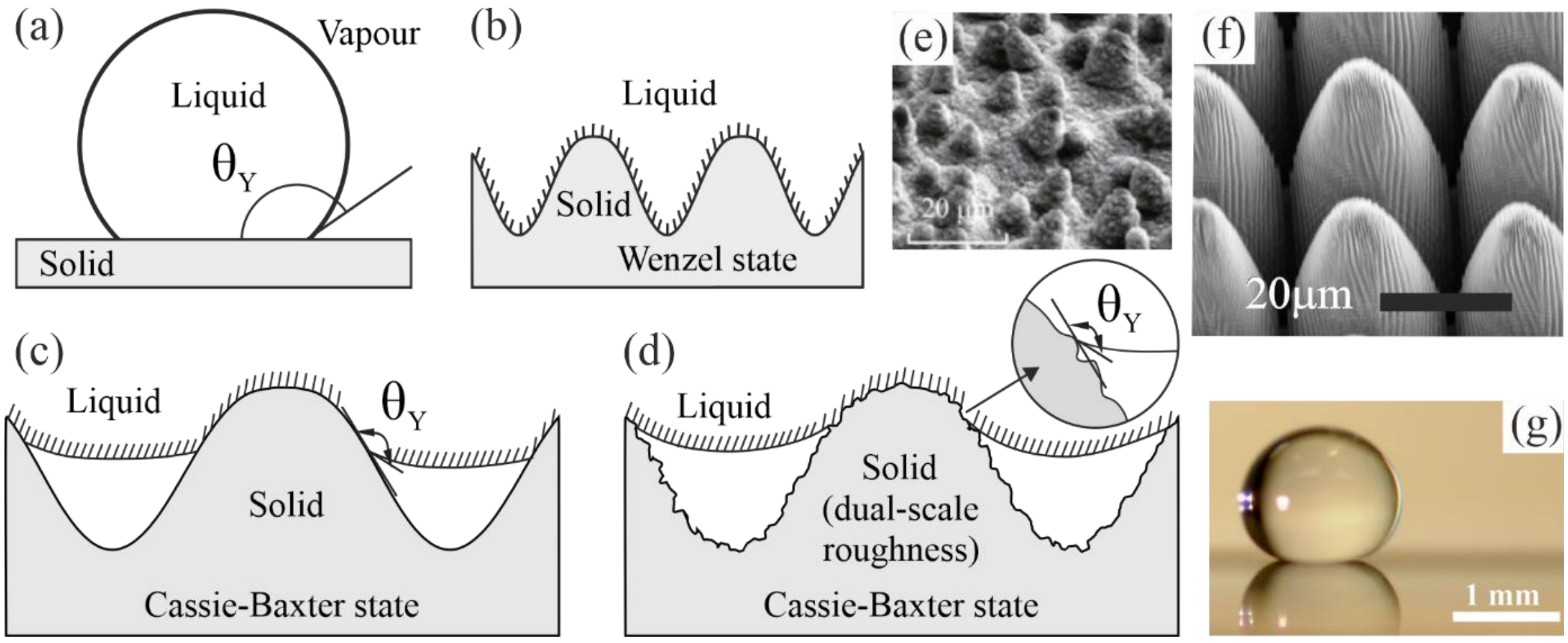

Fig. 10: (a) Definition of contact angle $\theta_Y$ [°] of a liquid (droplet) resting on a flat solid in a gaseous environment (solid–liquid–vapour system). (b) Wenzel state, in which the liquid fully wets the surface (c) Cassie-Baxter state, in which air is trapped below the liquid. (d) Liquid on a hierarchical (here dual-scale) surface topology composed of micrometer and nanometer scale surface roughness, resulting in increased Apparent Contact Angle

(APCA). (e) SEM micrograph of the surface topology of a Lotus leaf, composed of nanometer sized "hairs" on top of micrometer sized "pillars", showing hydrophobic and self-cleaning properties. (f) An SEM micrograph of a laser-processed surface resulting in a dual-scale surface texture — i.e. microscale pillars superimposed with LIPSS, replicating the surface topography of a Lotus leaf. (g) A water drop, with a large APCA, sitting on top of a laser dual-scale surface texture. Images (a)-(d) and (f),(g) adapted and reprinted by permission of [Arnaldo del Cerro, 2014]. Image (e) adapted from [Barthlott, 2017], Barthlott et al., Plant Surfaces: Structures and Functions for Biomimetic Innovations, Nano-Micro Lett. **9**:23, Copyright 2016 under Creative Commons BY 4.0 license. https://doi.org/10.1007/s40820-016-0125-1.

If the liquid is water, a (flat) surface is referred to as hydrophilic if $\theta_Y < 90°$ and hydrophobic—i.e. water repellent, if $\theta_Y > 90°$. A surface is considered superhydrophobic if $\theta_Y > 150°$. If a surface repels oily liquids it is referred to as oleophobic [Tuteja, 2007] and omniphobic if it repels all types of liquids. A superhydrophobic surface is usually only achieved if a roughness (texture) is present on the surface. In the latter case, $\theta_Y$ cannot be calculated using the above equation, but still the contact angle can be measured and is then referred to as the *Apparent Contact Angle* (APCA) [Stratakis, 2020]. Depending on the wetting state (or wetting regime), the APCA can be calculated using one of two models. If the liquid (droplet) is in full contact with the surface of the solid, the Wenzel model applies [Wenzel, 1936], see Fig. 10(b). If the liquid does not fully penetrate into the "valleys" of the surface roughness, but rests on the peaks of the texture, the Cassie-Baxter model applies [Cassie, Baxter, 1994], see Fig. 10(c). In this figure, the location of the contact line between the liquid and the solid is determined by the slope of the texture and the local Young's contact angle on the slope, see Fig. 10(c). The steeper the slope, the higher the contact line (closer to the peak) and therefore the less the liquid-solid contact area. As a result, a droplet easily rolls off, when the surface is tilted. Surface properties like self-cleaning, anti-icing and microfluidic applications are attributed to this small liquid-solid contact area. In the case of water, effects of increased APCA, when compared to flat surfaces, occur at micrometer scale roughness levels. If on this roughness scale, an additional nanometer scale roughness is superimposed, the APCA can be even further increased, due to the increased local slopes of the nanometer roughness, see Fig. 10(d). The self-cleaning properties of the Lotus leaf are attributed to this dual roughness scale, see Fig. 10(e). When designing a surface topology for improved wetting applications, LIPSS can be exploited on top of a micrometer scale texture, to obtain a dual-scale roughness, see Fig. 10(f)-(g). Many industrial processes and applications of LIPSS to adapt surface wetting of a surface or for liquid management are inspired by engineered replication of biological systems, such as the Lotus leaf [Stratakis, 2020 / Yong, 2015 / Nishimoto, 2013]. Also, many biomedical applications of LIPSS rely (partly) on surface wetting, see Section 2.3 *Biological Applications*.

## Self-Cleaning

Contaminants and debris on the surface of a product, such as tiles, textiles and solar-panels, can negatively influence not only its main function, but can also induce other effects such as tear and wear. It should be noted that, for optical applications, such as solar-panels or window glass, (super)hydrophobicity and transparency are competitive properties, because for roughness levels exceeding about 100 nm, optical transparency is compromised due to light scattering [Nishimoto, 2013]. A self-cleaning surface will reduce time and, therefore, costs associated with active cleaning procedures. If a surface exhibits (super)hydrophobic behaviour, water droplets easily roll off the surface and will pick up and carry away surface contaminants. Several authors, including Jagdheesh et al. and Piccolo et al. found that LIPSS increase the (apparent) contact angle on various materials up to tens of percent, when compared to "flat" or micrometer textured surface morphologies [Jagdheesh, 2011 / Piccolo, 2020]. Milles et al. [Milles, 2020], characterized the self-cleaning property of superhydrophobic hierarchical surface textures on aluminium, created by direct laser writing (DLW, using λ = 1064 nm and 8 ns and 14 ns pulse durations), direct laser interference patterning (DLIP) and LIPSS (using λ = 1064 nm and 10 ps pulse duration), which were contaminated with various particles including polyamide (PA) particles (~100 μm), see Fig. 11. Several droplets of 10 μl deionized water were dropped on contaminated laser-structured samples, which were tilted to allow droplet to roll of the sample and pick-up the particles, see Fig. 11(b). After each droplet, the number of particles on the region of interest was measured, see Fig. 11(c). The authors found high self-cleaning efficiencies of the DLW + DLIP + LIPSS structures with a remaining contamination as low as 1% when 100 μm diameter $MnO_2$ and PA particles were used, see Fig. 11(d).

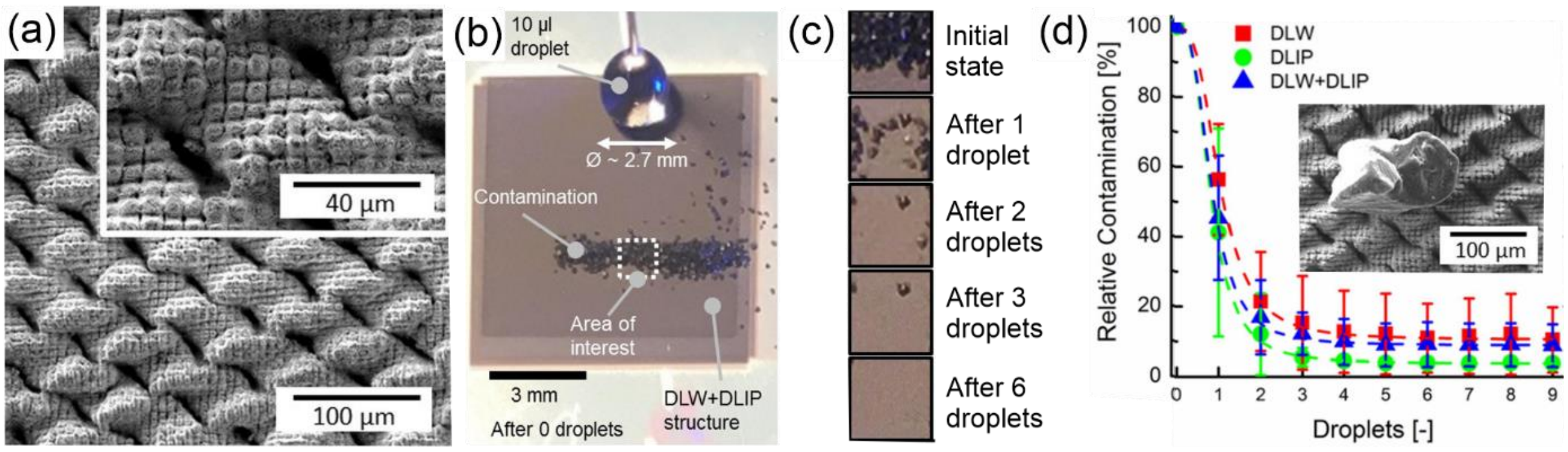

Fig. 11: (a) Top-view SEM micrograph of a hierarchical surface texture from resulting of the combination of DLW and DLIP and LIPSS showing self-cleaning properties. (b) Test setup used to assess self-cleaning properties of various surface textures. (c) The amount of particles on the region of interest was measured after each droplet rolled off. (d) Relative amount of particles in the various textures as functions of the number of rolled off droplets. (Reprinted and adapted from [Milles, 2020], Appl. Surf. Sci., Vol. **525**, Milles et al., Characterization of self-cleaning properties on superhydrophobic aluminum surfaces fabricated by direct laser writing and direct laser interference patterning, 146518, Copyright (2020), with permission from Elsevier).

## Fluidic Transport

In nature, hierarchical surface textures on the mm to nm scale are known to allow direct liquid flow. For example, the skin texture of moisture harvesting lizards living in arid regions consists of microstructured overlapping scales with capillary channels in between [Hermens, 2017]. The (micro)structure (10 mm to 30 μm) on the lizard scale "holds" a water film through which the skin can become super wettable. Due to the combination of capillary channels and the micro-structured scales, liquid is collected, as well as transported to the snout of the animal. Such surface properties can be exploited industrially for e.g. transport of cooling lubricants [Comanns, 2016] or lubricants in tribological applications, see Section 2.4 *Tribological Applications.* Hermens et al. [Hermens, 2017] employed a ps pulsed laser source at 532 nm to texture 16MnCr5 steel surfaces in order to mimick the scales and capillary channels of a desert horned lizard, see Figs. 12(a) and 12(b). The unidirectional transport of liquid, over distances in the range of centimeters, in/on such a surface texture is attributed to the network of interconnected capillaries that narrow and widen periodically. The authors characterized the liquid transport properties by depositing droplets of cooling lubricant K-40 at one side of the surface and observe the liquid transport using a video camera. It was found that the liquid propagates, from left to right in Fig. 12(c), at 0.49 mm/s velocity.

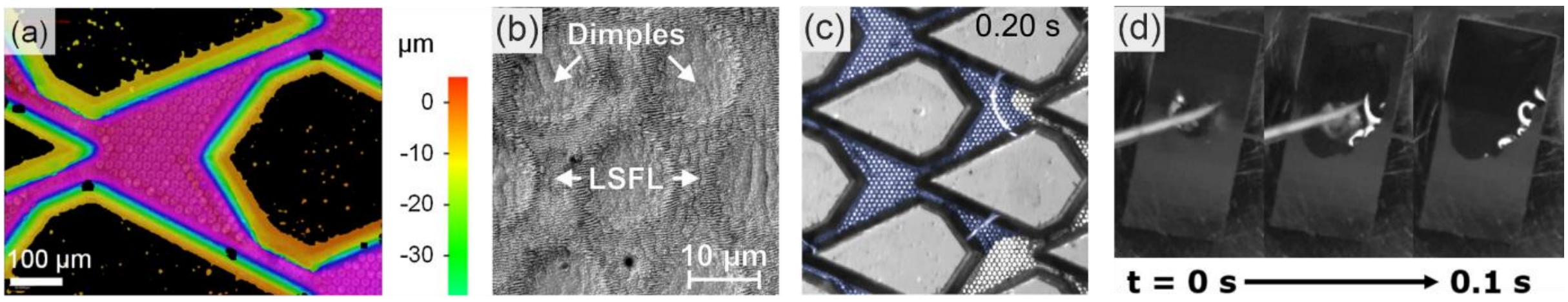

Fig. 12: (a) 3D surface topography of micrometer-sized 6-sided textured polygons ("scales"), with in between capillary "channels" consisting of micrometer-sized dimples (diameter ~10 µm) covered by sub-micrometer LSFL. (b) SEM micrograph (detail) of the surface texture in the capillary channels. (c) Image of lubricant K-40 (artificially coloured blue) at 0.2 seconds after a droplet was deposited near the left side of the image. (d) Sequence photographs (top-view) demonstrating the unidirectional flow of a water droplet, placed in the centre of a gradient of surface textures (top: Spikes, bottom: LSFL) on steel. (Images (a)-(c) reprinted from [Hermens, 2017], Appl. Surf. Sci., Vol. **418**, Hermens et al., Mimicking lizard-like surface structures upon ultrashort laser pulse irradiation of inorganic materials, 499–507, Copyright (2016), with permission from Elsevier. Image (d) taken from [Kirner, 2017b]. (Reprinted by permission from Springer-Verlag GmbH: Applied Physics A **123**:754 (Mimicking bug-like surface structures and their fluid transport produced by ultrashort laser pulse irradiation of steel, Kirner, S.V. et al.), Copyright (2017)).

As another example, Kirner et al. applied a 170 fs pulsed laser source at λ = 1026 nm to texture the surface of 42CrM04 steel with a gradient of surface morphologies ranging from LSFL over grooves to spikes (see Section 1.1 *Zoology of LIPSS*), by varying the laser fluence from 0.07 J/cm$^2$ (bottom of the rectangle) to 1.05 J/cm$^2$ (top of the rectangle), in order to mimic the transport properties of the skin of a bark bug (*Aradidae*) [Kirner, 2017b]. It was found, using video recordings, that, when applying an 8 µl droplet of deionized water in the center of the textured surface, shortly after laser processing, the liquid quickly spread at ~75 mm/s towards the area covered by spikes, and not to the area covered by grooves or LSFL, see Fig. 12(d).

Another facet related to the transport of fluids is the laser-based creation of *superwicking* surfaces. For this functionality, the combination of capillary forces, e.g., in laser-generated parallel microgrooves, along with the philicity of the nanostructure-covered inner channel surfaces can effectively transport liquids even against gravity [Vorobyev, 2009b], see Fig. 13.

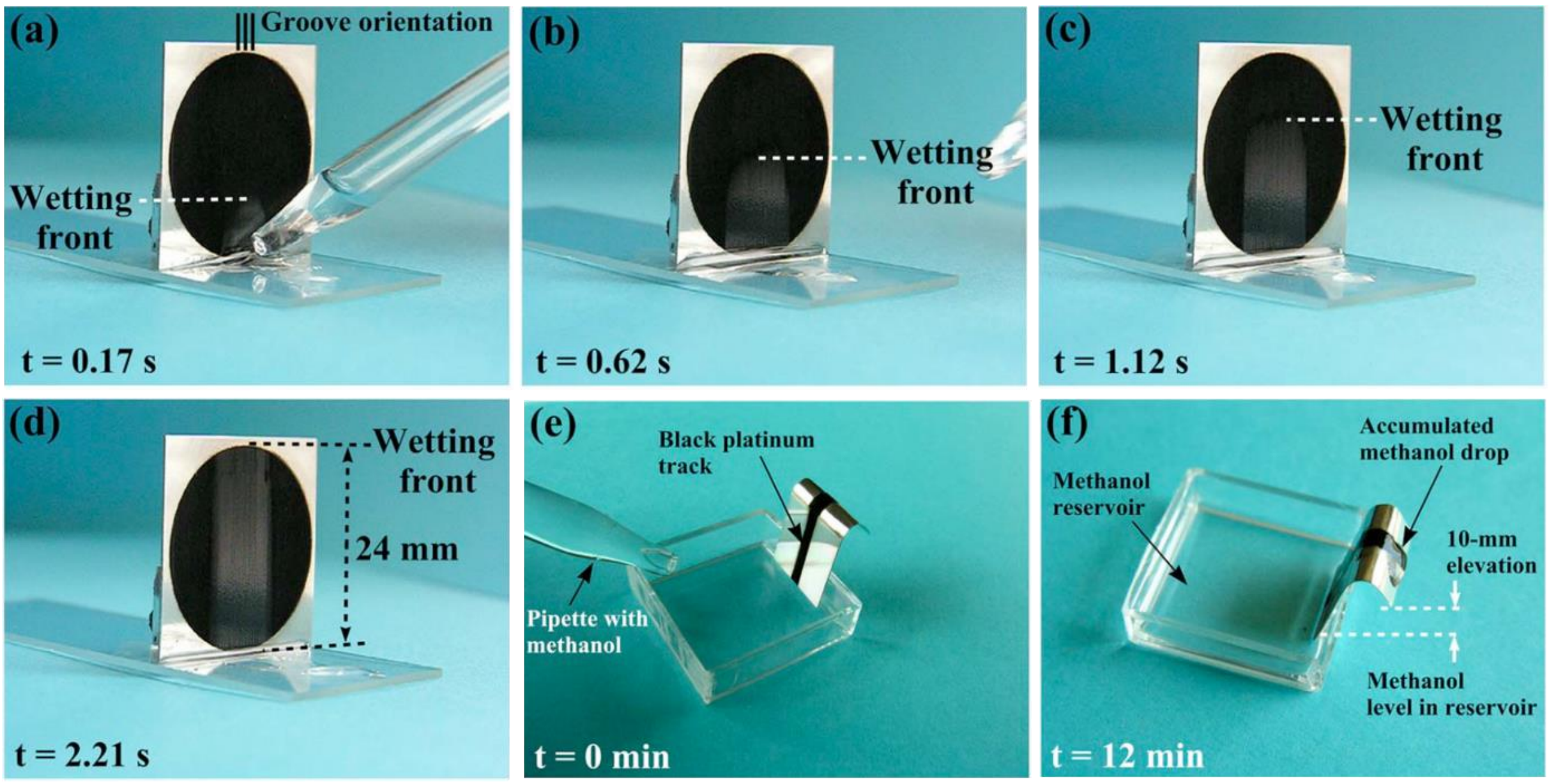

Fig. 13: (a)-(d) Photographs showing the methanol transport against gravity on a vertically standing laser-processed black platinum surface. (e)-(f) Photographs showing the vertical transport of methanol and liquid

accumulation across a sample edge. (Reprinted from [Vorobyev, 2009b], Vorobyev et al., Metal pumps liquid uphill, Appl. Phys. Lett. **94**:224102 (2009)], with the permission of AIP Publishing.)

Different applications based on such superwicking surface functionality were proposed and demonstrated already. Due to the capillary action, a liquid drop quickly spreads on the wicking surface and forms a thin film over a large surface area, finally resulting in efficient evaporation. Taking benefit of such a transport of water against gravity and the simultaneously strongly enhanced absorption of laser-processed black metals for optical radiation in the visible and infrared spectral range (see Section 2.1 *Optical Applications*), Singh et al. [Singh, 2020] demonstrated a solar-trackable superwicking black metal panel for photothermal water sanitation and desalination (see Fig. 14).

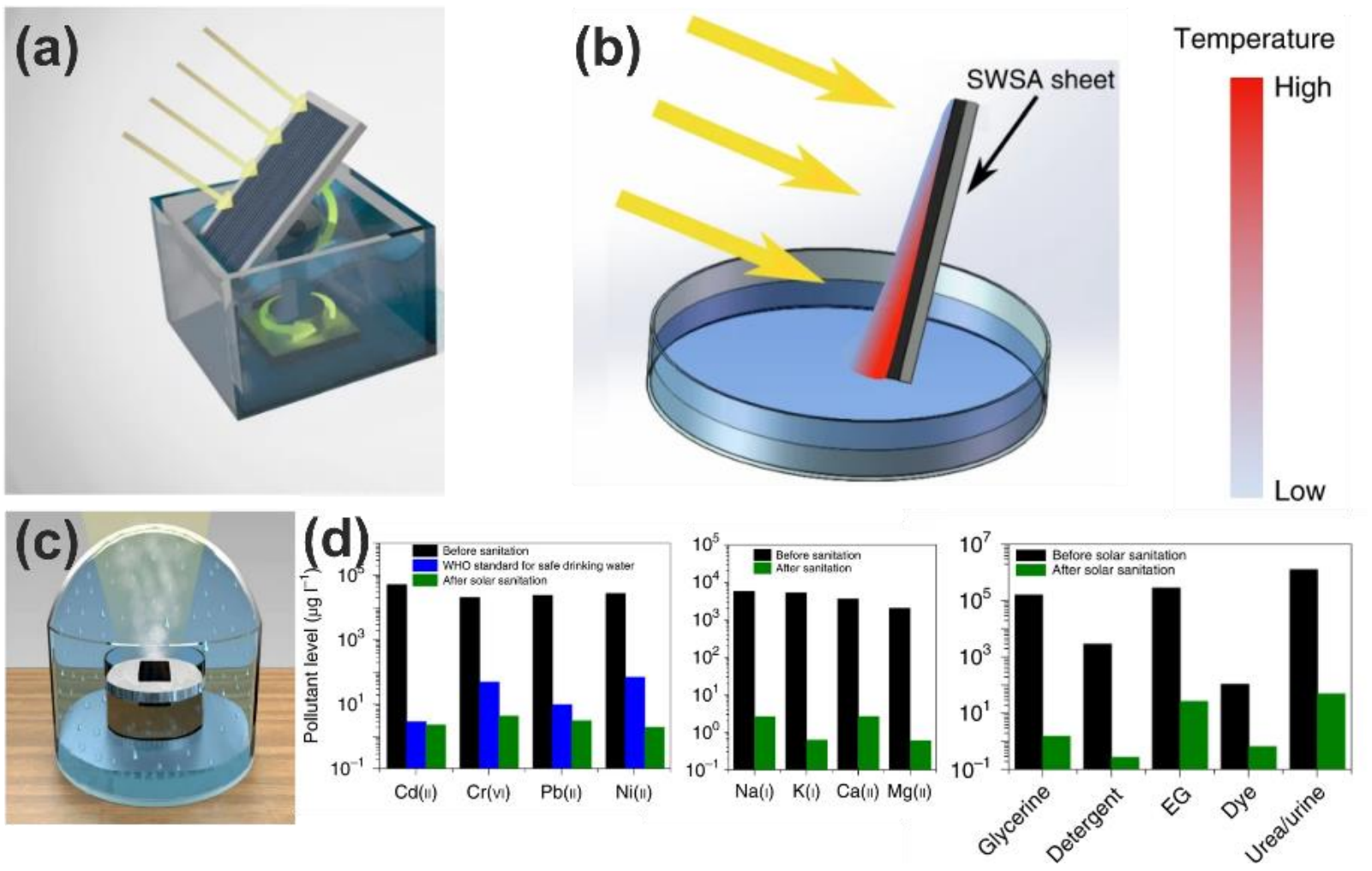

Fig. 14: (a) Schematics of a sun-tracking superwicking and super-light-absorbing (SWSA) aluminium solar-based water sanitizer that (b) promotes interfacial heating while ensuring that water is continuously transported across the solar absorber. (c) Schematic of the water sanitation/desalination setup. (d) Water sanitation of various heavy metals, salts, and industrial/domestic/human water pollutants (EG: ethylene glycol). (Reprinted (adapted) from [Singh, 2020], Singh et al., Solar-trackable super-wicking black metal panel for photothermal water sanitation. *Nat. Sustain*. **3**:938–946, Copyright 2020 under Creative Commons BY 4.0 license. Retrieved from https://doi.org/10.1038/s41893-020-0566-x).

On laser-processed aluminum, such superwicking surface functionalities even persist for water at elevated temperatures up to 120°C and may have applications in boilers with enhanced critical heat flux or in heat exchangers for enhancing energy efficiency in power generation [Vorobyev, 2021].

## Anti-adhesive Packaging for Liquid Food

Viscous liquids, which adhere to packaging surfaces of, for example, food, cosmetics and agro-chemical products leads to wastage, an increase of recycling costs, and consumer annoyance. Inspired by the *Nepenthes* pitcher plant, Karkantonis et al. created lubricant-impregnated textured surfaces (LIS) to improve the shedding behaviour of polymer surfaces of viscous liquids [Karkantonis, 2020]. In this approach, the lubricant is chosen such that it wets the micro/nanotextured surface instead of repelling it, and such that the viscous liquid, which is to be repelled, is immiscible with the lubricant. A 310 fs pulsed laser source ($\lambda$ = 1030 nm) was applied to texture an X6Cr17 stainless steel surface with a dual-scale texture consisting of a micrometric hatch pattern and LIPSS, see Fig. 15(a). This surface morphology was subsequently replicated in sheets of polystyrene (PS) and polypropylene (PP) by hot embossing. Silicone oil was selected as the lubricant.

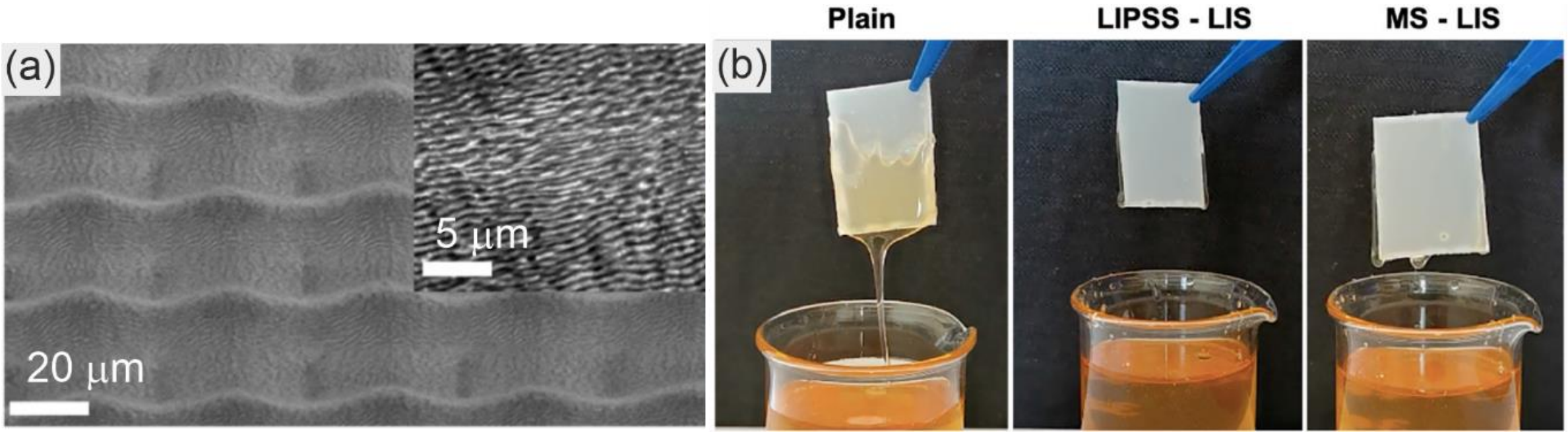

Fig. 15: (a) SEM micrograph of dual-scale laser textured steel surface, consisting of a micrometric hatch pattern and LIPSS (insert). (b) Photographs of the dripping behaviour of PP samples (impregnated with silicone oil) after submergence in honey; left: non-textured plain sample; center: sample covered with LIPSS; right: sample with dual-scale roughness. (Images adapted and reprinted from [Karkantonis, 2020], Surf. Coat. Technol., Vol. **399**, Karkantonis et al., Femtosecond laser-induced sub-micron and multi-scale topographies for durable lubricant impregnated surfaces for food packaging applications, 126166, Copyright (2020), with permission from Elsevier).

The shedding behaviour of various liquids was assessed by exposing the LIS polymer samples in various liquids, including honey, milk and ketchup. Figure 15(b) shows photographs of a submerging test of polypropylene (PP) samples in honey. The authors found that LIPSS-LIS treated thermoplastics demonstrated anti-adhesive properties even after 50 cycles with all tested liquids, while there was a slight stickiness on the dual-scale-LIS substrates after 43 and 20 dipping cycles in milk and honey, respectively.

## Drag Reduction

Drag is a force acting on a (solid) object when it moves in a fluid and acts in opposite direction to the relative motion of to object in the fluid. Drag reduction is beneficial in, for example, biomedical devices (e.g. in lab-on-a-chip applications), swim-suits, piping to transport liquids, ship hulls and aerospace industry [Nishimoto, 2013]. Compared to flat surfaces, reduction of drag can be achieved by surface textures, because a shear free boundary condition occurs at the air-liquid menisci, in the Cassie-Baxter state [Jagdheesh, 2010], see Figure 10(c)-(d) and the no-slip condition between the liquid and the roughness peaks of texture [Ahmmed, 2016]. Hence, reducing relative fraction of the (textured) solid surface that is in contact with the liquid, will result in drag reduction. Drag reduction can be quantified by measuring the so-called “slip length” of the (textured) solid, e.g. using a rheometer or Particle Image Velocimetry (PIV). A large (effective) slip length reduction, implies a large drag reduction. Therefore, Tanvi Ahmmed and Kietzig applied a <85 fs pulsed 800 nm laser source to texture the surface of pure copper with various micrometer scale patterns, including parallel lines (grate structures, see Fig. 16(a)) and various pillar structures (square and rhombic shaped, see Figs. 16(b)-(c)) all of which the sidewalls were found to be covered with LIPSS and the peaks with nanoparticles [Ahmmed, 2016].

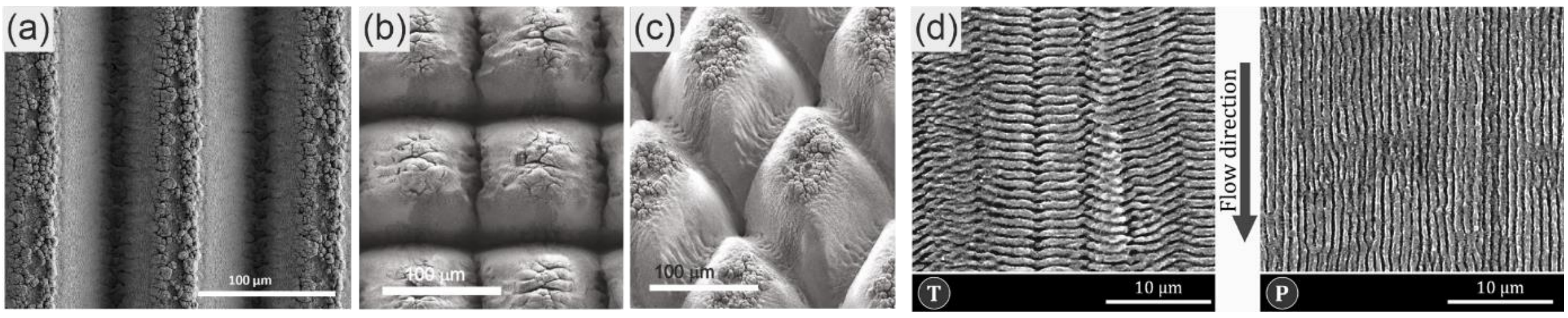

Fig. 16: (a)-(c) SEM micrographs of various multiscale surface morphologies for liquid drag reduction, consisting of micrometer scaled patterns superimposed by LIPSS on the slide walls and nanoparticles on the peaks. (d) SEM micrographs of LSFL on mould steel aiming a drag reduction in polymer injection moulding. (Images (a)-(c) republished with permission of Royal Society of Chemistry, from [Ahmmed, 2016], Drag reduction on laser-patterned hierarchical superhydrophobic surfaces, Ahmmed, K.M.T. and Kietzig. A.-M., Soft Matter **12**:4912–4922, 2016]; permission conveyed through Copyright Clearance Center, Inc. Image (d) adapted and reprinted from [Sorgato, 2018], CIRP Annals, Vol. **67**, Sorgato et al., Effect of different laser-induced periodic surface structures on polymer slip in PET injection moulding, 575–578, Copyright (2018), with permission from CIRP).

Next, the samples were coated with fluoroalkylsilane in order to chemically enhance the hydrophobicity of the surface. The authors employed water and a rheometer to quantify the slip lengths, by calculating from measured torques at different shear rates. The slip length on a smooth (so untextured) hydrophobic surface is small and varies from nm to a few µm [Ahmmed, 2016]. The authors found significantly larger slip lengths of their samples, varying from 70 µm to about 140 µm.

As another example, Sorgato et al. employed an 8 ps, 1064 nm laser source to cover mould steel with LSFL (see Fig. 16(d)) in order to reduce the drag of a polymer (PET) in an injection moulding setup [Sorgato, 2018]. That is, because especially in injection moulding of small plastic parts, e.g. in electronic packaging applications, drag is high. Therefore, high pressures are required to fill the mould. The authors covered mould insert with two different orthogonal orientations of LSFL, parallel (P) and transverse (T) to the flow direction of the polymer, see Fig. 16(d). It was found that LSFL transverse to the melt flow have a negligible effect on the injection pressure, whereas LSFL parallel to the flow direction was found to reduce the required injection pressure up to 23%.

## Leidenfrost Effect

If a liquid droplet is applied on a surface which is significantly hotter than the boiling point of the liquid, an insulating vapour “cushion” forms below the droplet, such that it hovers over the solid and does evaporate slowly. This phenomenom is known as the Leidenfrost effect. By applying a surface texture the Leidenfrost temperature — i.e. the critical temperature of the solid over which the Leidfrost effect occurs — reduces [Dubnack, 2020 / Arnaldo del Cerro, 2014 / Marín, 2012]. The roughness level required to reduce the Leidenfrost temperature for water is in the µm range [Arnaldo del Cerro, 2014 / Marín, 2012]. Moreover, based on the Leidenfrost effect, assymetric surface textures, such as a saw tooth profile, can be exploited to direct the movement of droplets, even against gravity, similar to the superwicking effect (see above). These movements are induced by pressure gradients introduced by the asymetric inclination of the surface morpholgy, as well as due to rotation of droplets due to Marangoni flows. These effects, and the reduction of the Leidenfrost temperature have potential applications in e.g. *lab-on-a-chip* devices and piping to transport liquids, as it reduces drag. Dubnack et al. employed a 300 fs, 1025 nm wavelength laser source to cover the tips of a saw tooth profile in aluminium with LSFL, see Fig. 17(a) [Dubnack, 2020]. The samples were coated with fluorooctyltriethoxysilane to increase the hydrophobicity chemically. Next, the authors measured the temperature which initiated a water droplet (volumes 60 to 200 µl) to start moving. This temperature was found to equal about 170 °C for the texture without LIPSS and only about 115 °C for the textures with LIPSS, see Fig. 17(c).

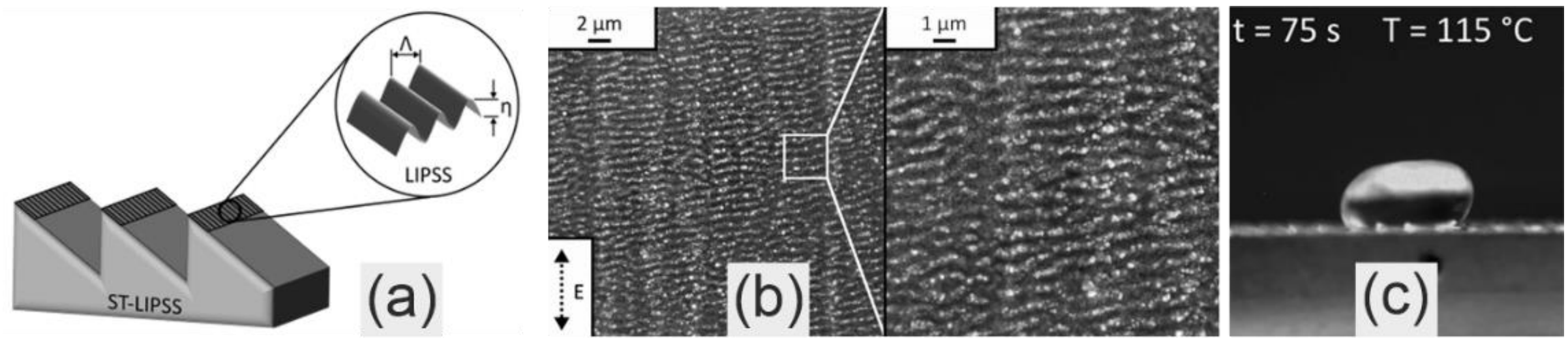

Fig. 17: (a) Millimetric saw tooth surface profile covered with LIPSS to accelerate water based on the Leidenfrost effect. (b) SEM micrograph of the LSFL on the tips of the saw tooth surface profile on the aluminium substrate. (c) Frame from a video sequence showing a moving droplet (starting from the left of the frame) at a reduced Leidenfrost temperature. (Reprinted from [Dubnack, 2020], Appl. Surf. Sci., Vol. **532**, Dubnack et al., Laser-induced Leidenfrost surfaces, 147407, Copyright (2020), with permission from Elsevier).

## Anti-Icing

Ice (frost, glaze, rime, snow) can cause problems when it adheres and accumulated on solar panels, aircraft, wind turbines, power- or communication-lines, to name a few [Volpe, 2020]. Anti-icing properties can be obtained by increasing the surface roughness of a substrate in order to increase the hydrophobicity of the surface. Designing a (super)hydrophobic surface aims at increasing the time delay in the solidification of water, promote roll-off of water before freezing occurs (see self-cleaning) and reduces adhesion of ice after it formed on the surface [Volpe, 2020]. A reduction of the relative solid fraction of the surface in contact with the water droplet(s) promotes anti-icing properties. A low relative solid fraction is achieved in the Cassie-Baxter state (see Figs. 10(c) and 10(d)), which lowers the heat conduction between the droplet and the surface, promotes roll-off and reduces ice adhesion, when compared to the Wenzel state and to flat or smooth surfaces.

Milles et al. assessed the freezing time of water droplets on aluminium surfaces covered with hierarchical surface textures created using the same techniques (DLW and DLIP) and laser-systems as described above (see self-cleaning) [Milles, 2009]. The freezing time of the DLW structures, DLIP structures and the hierarchical structures (DLW+DLIP+LSFL, see Fig. 11(a)) was measured by imaging the solidification of 8 µl water droplets on the surface in air of -20 °C, see Fig. 18(a). As can be observed from this graph, the freezing time of the droplet on the hierarchical structure (22.3 s) is about 2.6 times longer than on the smooth reference sample (8.7 s). Furthermore, the authors found that 13 µl droplets, when dropped from a height of 20 mm on samples cooled down to -20 °C, would only bounce on the hierarchical structures, but not on the DLW structures, nor the DLIP structures. Droplet rebounding was found to increase the freezing time.

Römer et al. studied the roll-off of water droplets on a laser textured stainless steel sample in a representative rain test and a rime test [Römer, 2009]. To that end, the authors employed a 7 ps, 1030 nm wavelength laser to create a hierarchical surface texture consisting of pillars (height 20 µm, pitch 20 µm) covered with LSFL, see Fig. 18(b). The pitch of the pillars was chosen such that several peaks will support a water droplet in the Cassie-Baxter state, see Fig. 10(d). Next the surface was covered by a Perfluorinated OctylTrichloro-Silane (FOTS) coating to increase the hydrophobicity of the surface chemically. After the rain test (water droplets at -0 °C, climate chamber at -5 °C, 9 m/s wind speed, which are icy conditions for airplanes) of 10 seconds a photograph was taken, see Fig. 18(c). It was found that the laser textured area is almost completely free of frozen droplets. Further, it was found, in a rime ice test, that the laser textured surface only slightly slows down the formation of rime, see Fig. 18(d). The latter can be attributed to the fact that the diameter of water droplets forming rime are (much) smaller than the pitch of the pillars in Fig. 18(b).

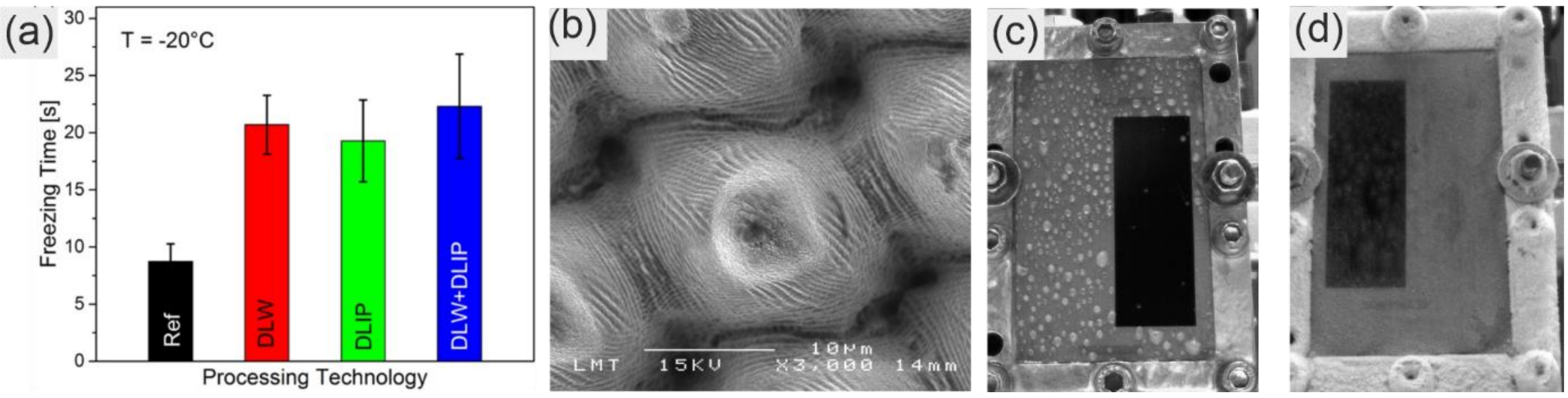

Fig. 18: (a) Average freezing time of water droplets on an untreated reference, DLW, DLIP, DLW + DLIP+LSFL structures. (b) Top-view SEM micrograph of a steel surface texture consisting of 20 µm pitched pillars covered with LSFL (c) Photograph of the FOTS coated laser textured steel sample, 10 seconds after the rain test. The laser textured area (black) of several cm$^2$ in size is almost completely free of frozen droplets. (d) Photograph of the FOTS coated laser textured steel sample after the rime test. (Image (a) reprinted from [Milles, 2019], Milles et al., Fabrication of superhydrophobic and ice-repellent surfaces on pure aluminum using single and multiscaled periodic textures, Sci. Rep. **9**:13944, Copyright 2019 under Creative Commons BY 4.0 license. Retrieved from https://doi.org/10.1038/s41598-019-49615-x. (Images (b)-(d) adapted and reproduced from [Römer, 2009], Römer et al., Ultra short pulse laser generated surface textures for anti-ice applications in aviation. Proceedings International Congress on Applications of Lasers & Electro-Optics (ICALEO), 2-5 Nov., Orlando, USA, p. 30–37 (2009), https://doi.org/10.2351/1.5061570 with the permission of the Laser Institute of America.)

## 2.3 Biological Applications

In this Section biological applications of laser-induced surface structures are addressed. It will be discussed that LIPSS can be used to control the adhesion and growth of bacterial films and affect the alignment of cells on medical implants. Moreover, LIPSS can be used in biochemistry for significantly enhancing Raman spectroscopic signals in lab-on-a-chip devices.

### Biofilm Growth Control

Most common issues with prosthetics, such as e.g., dental implants [Hanif, 2017], hip and knee prosthetics [Löwik, 2017], as well as biomedical devices such as catheters, cardiac pacemakers and stents [Siddiquie, 2020] are joint infections and thrombosis. This causes additional hospitalization and costs and may end lethal in the worst case. A solution to this problem could be the implementation of anti-bacterial surfaces on such devices.

The wings of butterflies and cicadas as well as the skins of geckos show anti-bacterial properties mainly due to well-ordered micro- and nanostructures [Tripathy, 2017]. LIPSS on different stainless steel alloys [Fadeeva, 2011 / Truong, 2012 / Cunha, 2016 / Chan, 2017 / Epperlein, 2017 / Lutey, 2018 / Zwahr, 2019 / Müller, 2020 / Selvamani, 2020 / Luo, 2020] and various types of polymers [Schwibbert, 2019 / Siddiquie, 2020] have shown similar antibacterial properties.
It is to be believed that the laser process leads to an oxidation of metal surfaces, which may act bactericidal [Chan, 2017 / Müller, 2020]. Additionally, it may be concluded that the presence of a topographic surface relief with characteristic dimensions smaller that the bacterium size and with very dense features reduces the contact area between bacteria and the surface, inhibiting bacterial colonization and reducing bacteria retention [Fadeeva, 2011 / Truong, 2012 / Cunha, 2016 / Chan, 2017 / Schwibbert, 2019 / Müller, 2020]. However, this depends on the sizes and shapes of the bacteria as shown in Fig. 19 by Epperlein et al. [Epperlein, 2017]. That figure shows fluorescence microscopy images of corrosion-resistant steel (V4A) samples partially covered with LIPSS (LSFL) with a period of about 700 nm, colonized for (a) 18 hours with *Escherichia coli* and (b) 20 hours with *Staphylococcus aureus*. It can be observed that the rod-shaped *E. coli* bacteria (with dimensions of ca. 0.5 µm × 2 µm) exhibit reduced adhesion on the LIPSS-covered area, whereas the spherically-shaped *S. aureus* bacteria (diameter 0.5 - 1 µm) show increased adhesion on the LIPSS-covered area. It was stated by the authors, that the increased adhesion results from the bacteria shape aligning with the direction of the nanostructures and hence increases the contact area between bacteria and surface [Epperlein, 2017].

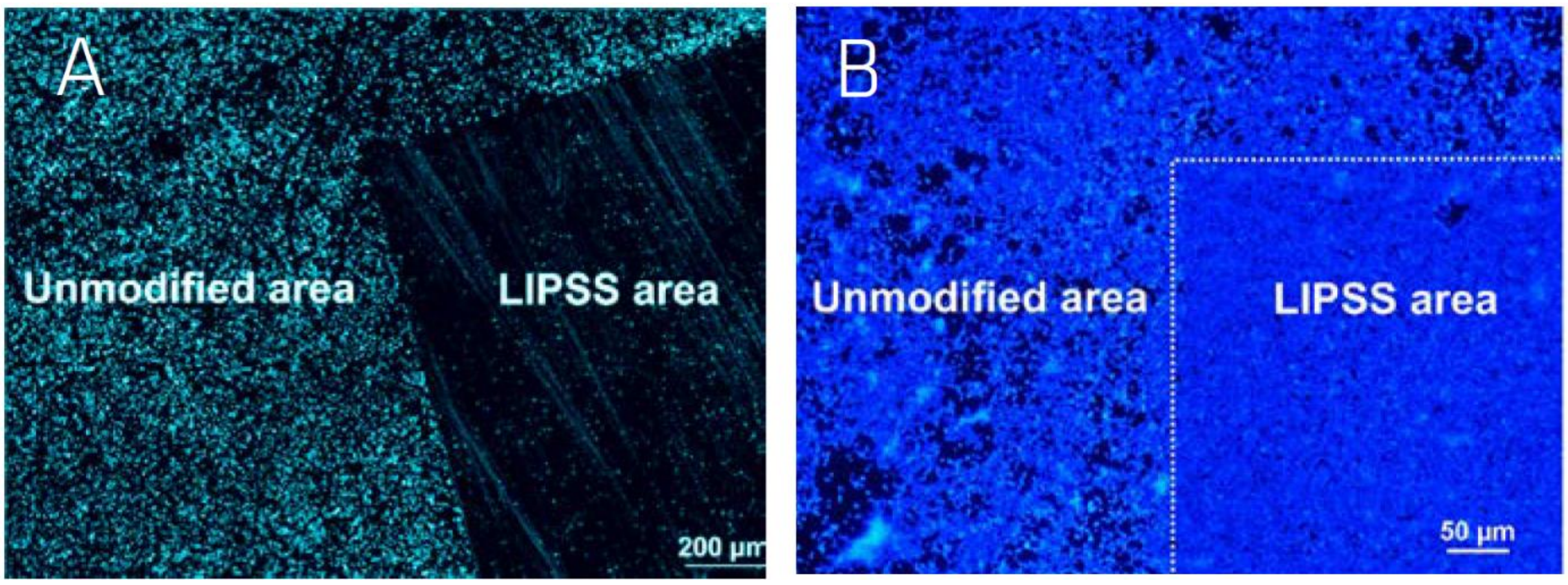

Fig. 19: Fluorescence microscope images of steel samples (V4A), partially processed with LSFL and subsequently colonized with dye-stained (a) *E. coli* bacteria after incubating of 18h and (b) *S. aureus* bacteria after incubating of 20h. (Reprinted from [Epperlein, 2017], Appl. Surf. Sci., Vol. **418**, Epperlein et al., Influence of femtosecond laser produced nanostructures on biofilm growth on steel, 420–424, Copyright (2017), with permission from Elsevier).

Hierarchical micro-/nanostructures consisting of a combination of micro-sized structures covered with nano-scaled features can also act bactericidal, as was shown e.g. by Luo et al. [Luo, 2020]. Figure 20(b) shows *E. coli* on different types of laser-induced surface structures on titanium. That is, “type 1” consists of LSFL with a spatial period of about 400 nm, “type 2” contains LSFL interrupted by shallow sinusoidal grooves with a distance of about 1 µm and “type 3” is composed of LSFL with a period of about 500 nm interrupted by grooves which are about 200 nm deeper than the grooves of type 2. All three structure types show an anti-bacterial effect, as can be summarized in Fig. 20(a), showing the reduction of bacteria counts in percentage of the three surface structure types compared to polished titanium as a reference. All three types also show bactericidal effects, with structure type 3 showing the highest bacteria reduction by 56%. The proportion of live and dead bacteria after incubation time of the live and dead bacteria vary for the surface topographies. It can be observed in Fig. 20(b) that the bacteria tend to form clusters of several bacteria (highlighted by a yellow circle) on structure type 1 (LSFL) in order to prevent rupturing. Cluster formation was observed less on type 2 and 3 but the bacteria were located more at the bottom of the grooves, especially for type 3. Rupturing of the bacteria is caused either on top of nanostructures, which leads to a platy form or between two microstructures, leading to a lacerated form. For surface texture type 1 mostly platy ruptured dead bacteria were observed, whereas for structure type 3 the lacerated form is primarily present [Luo, 2020].

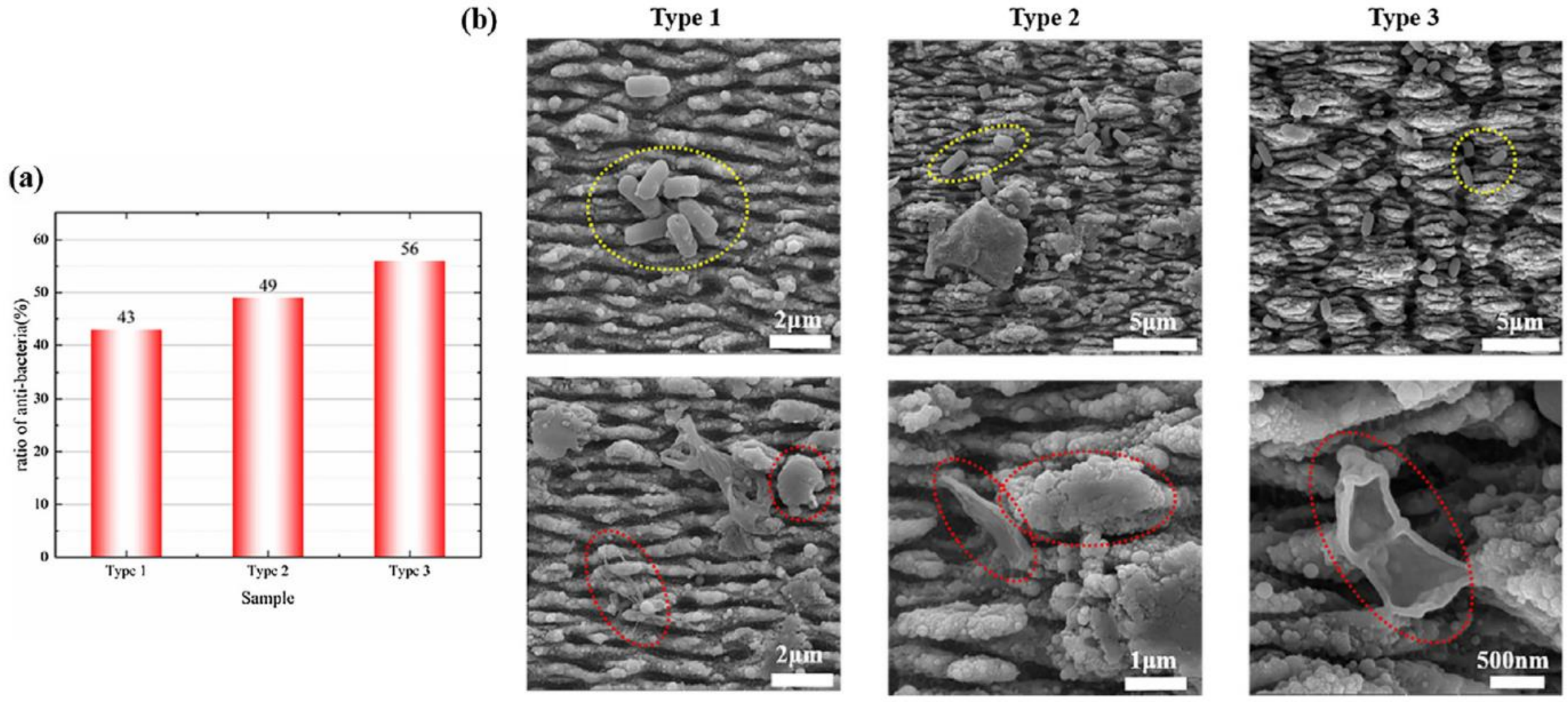

Fig. 20: (a) Anti-bacterial ratio (the total dead cells on LIPSS of the total cell numbers on polished titanium) for *E. coli* of three types of surface structures. (b) Top-view SEM images of bacterial cells adhesion states for three types of surface morphologies after 24h of bacterial cell culture. Yellow ellipses show survival strategies such as cluster formation over several LIPSS ridges or cells lying between the LIPSS maintaining their original shape. Red ellipses show two types of dead cells including platy form and lacerated form. The platy form means the dead cells spread wrapping the top of surface structures and the lacerated form represents the stretched dead cells hanging between the surface structures. (Reprinted from [Luo, 2020], Opt. Las. Technol., Vol. **124**, Luo et al., Biocompatible nano-ripples structured surfaces induced by femtosecond laser to rebel bacterial colonization and biofilm formation, 105973, Copyright (2019), with permission from Elsevier).

Conceptually, different surface roughness or nanoscale topography-mediated effects can be distinguished that may all led to a reduced bacterial colonization, for example

- Bactericidal effects, where sharp or nanometric spiky features of the surface topography can perforate the cell membrane [Elbourne, 2017];
- Obstruction or hampering of proliferation of bacterial cells, e.g., through their spatial confinement [Mahanta, 2021];
- Reduced adhesion of bacterial cells during the early stage of biofilm formation;
- etc.

In the latter context, surface topography-induced hydrophobicity can feature a low surface wettability and low surface energy to promote a low drag surface under flow conditions that is finally preventing microbial colonization at the surface through the reduced strength of adhesion of the bacteria to surface [Linklater, 2021].

Moreover, very recently, Richter et al. [Richter, 2021] demonstrated that a repellence of *E. Coli* bacteria of up to ~90% can be obtained by controlling the spatial period of UV ns-laser processed sub-micrometric LSFL on polyethylene terephthalate (PET) foils. This study pointed out that the efficiency of bacteria-repellent surfaces can be significantly improved when surface nanostructures impair the adhesion for bacterial nanofiber-appendages, such as pili. These pili typically feature diameters of less than 10 nm and are establishing bacterial cell-surface and cell-cell contacts as seen in the SEM micrographs shown in Fig. 21.

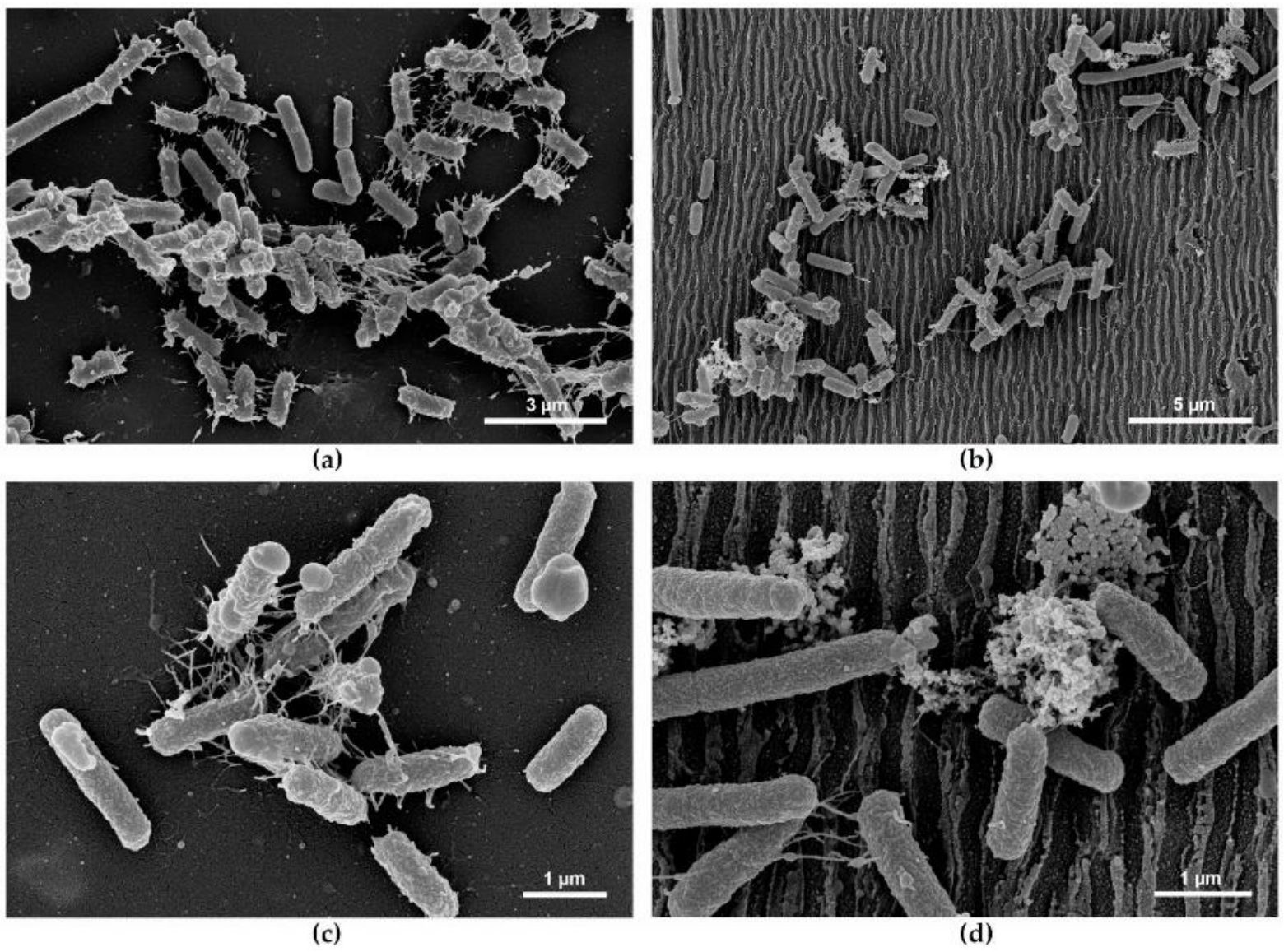

Fig. 21: SEM micrographs of *E. coli* bacteria on (a,c) non-irradiated (flat) PET and (b,d) laser-structured PET covered with LSFL of < 300 nm spatial period. Note the different scale bars. (Reprinted from [Richter, 2021], Richter et al., Spatial Period of Laser-Induced Surface Nanoripples on PET Determines *Escherichia coli* Repellence. *Nanomaterials* **11**:3000 (2021), Copyright 2021 under Creative Commons BY 4.0 license. Retrieved from https://doi.org/10.3390/nano11113000)

It must be noted that certain micro- and nanostructures are not well suited for metal-on-polymer bearing surfaces in joint prosthetics due to increased friction and abrasion, which results in a shortening of the lifetime of the joint prosthetics [van der Poel, 2019]. Nonetheless, for static implants, laser surface texturing can offer a valid technique to control biofilm growth on the implant surfaces. Moreover, in aqueous media, surface wetting effects of the laser-processed surfaces (see Section 2.2 *Liquid Management and Surface Wetting*) may be relevant for bacterial film formation.

## Cell Growth and Alignment

Another reason to add surface micro- and nanostructures on medical devices is to control the growth and alignment of biological tissue at the implant surface (see also **Chapter 31** (Buchberger et al.)).). While an accelerated and aligned cell growth for improved osseointegration is preferable for the interface between bone tissue and load bearing implants, such as dental implants or hip joint prosthetics [Li, 2020a], a decreased cell growth of fibroblasts is desired for other implants which are intended to be removed after some time, such as catheters, cardiac pacemakers or contraceptives [Heitz, 2017].

One of the standard materials for dental and orthopedic implants is Ti6Al4V titanium alloy because of its excellent mechanical properties such as low weight, high corrosion resistance and hardness, but is also a bioinert material which osseointegration abilities could be further improved to reduce the possibility of long-term implant failure [Li, 2020a]. Polystyrene is a widely used biomaterial for cell culture applications for its non-toxicity and low production cost [Rebollar, 2008 / Wang, 2008], but its hydrophobic surface hinders cell adherence and growth [Wang, 2008]. Therefore, the relatively fast, contact-less one-step LIPSS processing is a viable choice for enhancing biocompatibility via topographical and chemical changes due to the fact, that it can be applied on a wide range of materials, ranging from strong absorbing metals to transparent polymers [Heitz, 2017 / Petrovic, 2020 / Mezera, 2019].
It was shown that regular nanostructures (LSFL) on titanium alloys and polystyrene increase the growth rate, adhesion and density of mammalian cells [Rebollar, 2008], osteoblastic cells [Li, 2020a], glioma cells [Wang, 2008] and fibroblastic cells [Petrovic, 2020]. Moreover, the cells align parallel to the LSFL ridges and their shape elongates along the latter. Figure 22 by Rebollar et al. [Rebollar, 2008] exemplifies the elongating and parallel aligning of mammalian cells along LSFL processed by ns-UV laser radiation on polystyrene surfaces. This alignment occurred for LIPSS periods above a critical, cell type specific value.
In contrast, laser-generated hierarchical micrometric spikes covered by nanometric LSFL on titanium alloys and silicon wafers [Ranella, 2010] can suppress the growth of fibroblasts, particularly if an additional anodization step is performed after the laser processing [Heitz, 2017 / Fosodeder, 2020 / Lone, 2020]. It was shown by Klos et al. [Klos, 2020], that cell growth, adhesion and migration towards predetermined areas can be controlled by a site-selective combination of nanometer-scaled LSFL and micrometer-scaled spikes.

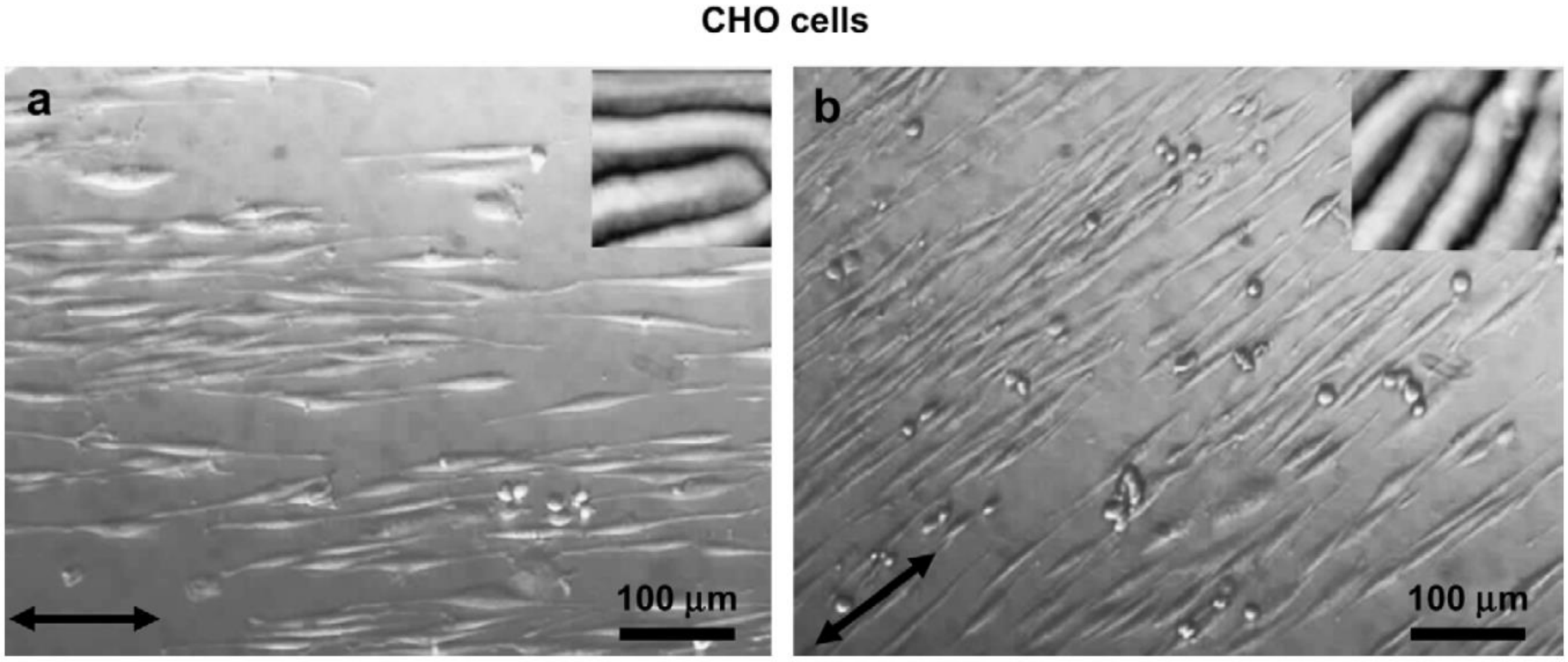

Fig. 22: Phase contrast optical micrographs of CHO-K1 cells (a) 27 h and (b) 52 h after seeding on PS foils laser-processed at an angle of incidence of $\theta = 45°$. Arrows indicate the direction of the LIPSS. In the insets typical magnified atomic force microscopy images (1.4 µm × 1.4 µm) of the LIPSS textured PS foils are shown. (Reprinted from [Rebollar, 2008], Biomaterials, Vol. **29**, Rebollar et al., Proliferation of aligned mammalian cells on laser-nanostructured polystyrene, 1796–1806, Copyright (2008), with permission from Elsevier).

## SERS Lab-on-a-chip

It was shown by several authors [Chang, 2011 / Buividas, 2012 / Li, 2020b] that LIPSS on suitable materials can be beneficial for surface enhanced Raman scattering (SERS) - leading to a sensitivity increasement of up to 2000 times compared with flat samples [Buividas, 2012]. The approach may be useful for applications in food safety, biosciences, and eco-pollution. Moreover, given its flexibility, laser processing is an outstanding candidate for SERS Lab-on-a-chip products due to its ability to structure complex shapes [Chang, 2011], its simple, precise and controllable process for achieving fluidic inlets, hydrophilic, liquid self-propagating microchannels and LIPSS using a single laser processing setup [Li, 2020b]. An example of such a laser processed SERS Lab-on-a-chip is shown in Fig. 23 [Li, 2020b], where (A) shows an overview over the whole chip, (B) shows a liquid inlet, (C) shows the microchannels, and (D) shows the LIPSS processed SERS active site.

This concept of spatially confining and transporting the chemical analyte through laser-textured surfaces can be even extended by combining it with the idea of evaporational concentration enhancement. This approach was realized by Hu et al. [Hu, 2020] via fs-laser processing of polytetrafluoroethylene surfaces to generate extremely superhydrophobic surfaces. For the spatial confinement of analyte droplets, small square-shaped areas of the surface remained unprocessed and served as “trapping areas” for the analyte. In combination with evaporation concentration of the analyte, e.g., rhodamine 6G dye molecules with a concentration of $10^{-6}$ M it was shown that the corresponding spectroscopic intensity detected via the SERS effect can be enhanced by up to two orders of magnitude on the laser fabricated samples when compared to the non-processed surface. Pavliuk et al. demonstrated an easy-to-implement laser-printed device for ultrasensitive SERS-based identification of various analytes dissolved in water droplets at trace concentrations sensitivity levels down to $10^{-12}$ M [Pavliuk, 2020].

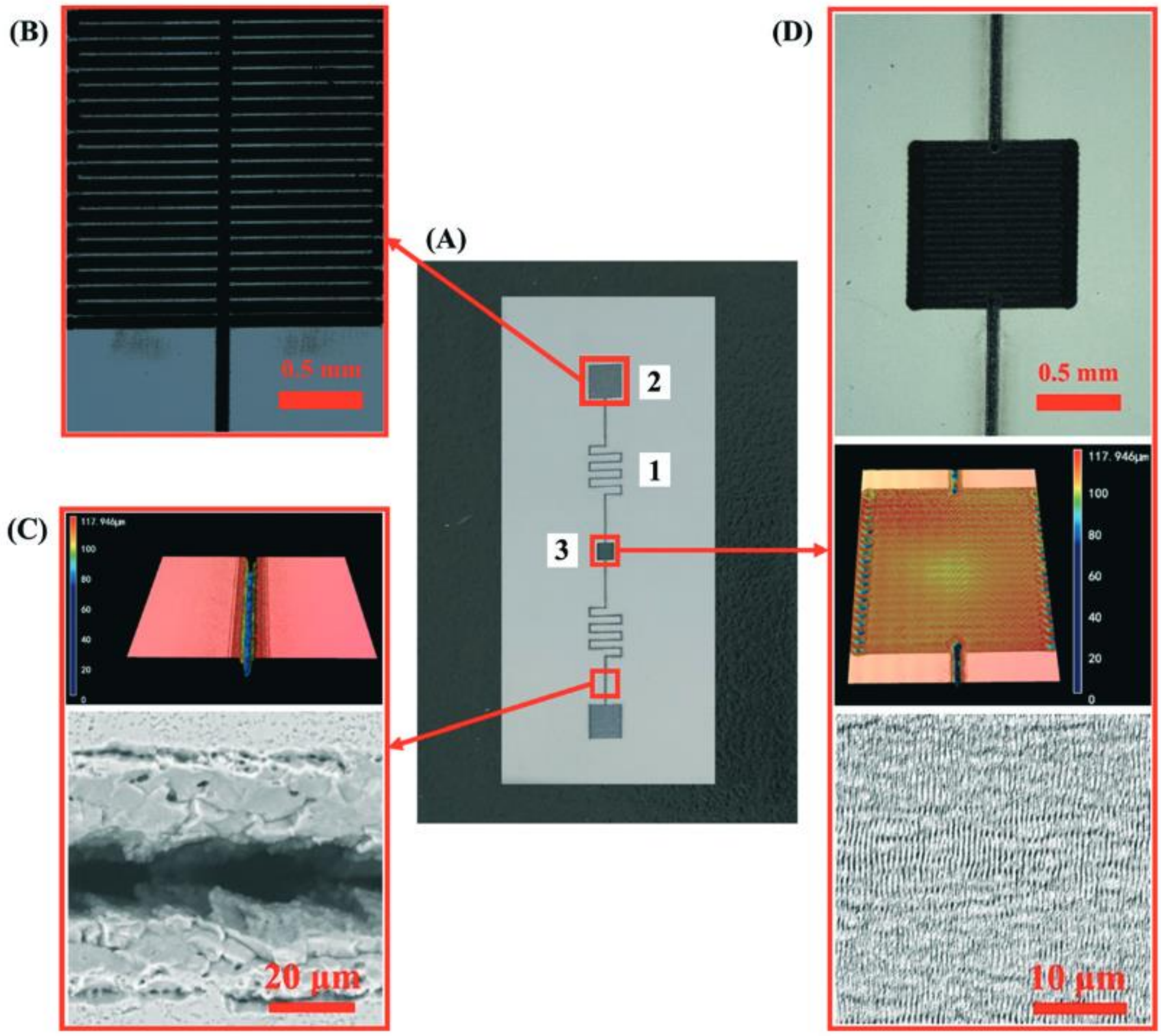

Fig. 23: (a) Photograph of the whole microfluidic platform. (b) Optical image of the 2 × 2 $mm^2$ square inlet. (c) 3D topography and SEM images of the microchannel. (d) Optical, 3D-topography and SEM images of the SERS active site. (Republished with permission of Royal Society of Chemistry, from [Li, 2020b], A self-driven microfluidic surface-enhanced Raman scattering device for $Hg^{2+}$ detection fabricated by femtosecond laser, Li, X., et al., Lab on a Chip **20**:414–423, 2020; permission conveyed through Copyright Clearance Center, Inc.).

## 2.4 Tribological Applications

This Section provides a brief survey on how LIPSS-covered surface can provide beneficial tribological properties, e.g., for the reduction of the coefficient of friction (CoF) or the wear between different solid bodies being in mechanical contact – or even for the manifestation of a passive directional transport of liquid lubricants towards the tribological contact area. Such a reduction of the CoF, even if nominally small, may have a tremendous potential for energy (fuel) saving and $CO_2$ reduction when accumulated over the entire lifetime of technical devices [Holmberg, 2017]. Simultaneously, the reduction of wear can significantly increase the servicing-intervals and enhance the devices life cycles. Hence, tribological applications of LIPSS and other tailored microstructures bear an enormous societal and economic potential.

### Reduction of Friction and Wear

It is a perception of paramount importance that properties such as the CoF (often referred to via the symbol $\mu$) or wear rates are not simply material properties but must be considered as system properties [Czichos, 2015]. A tribological system usually consists of (at least) two bodies being in touch with each other via their surfaces – either through a conformal or a non-conformal contact, featuring locally potentially very complex and even transient (multi-)contact areas. Moreover, some macroscopic forces are typically acting between the two bodies (e.g., shear forces or a normal force $F_N$), among other relevant environmental parameters, such as the temperature, humidity, etc. Additionally, there may be a liquid or solid lubricant added in the tribological contact region (e.g. graphite, grease, oil, water, etc.). Often the relative motion between the two bodies can be further characterized through their relative velocities $v_{rel,}$ some specific stroke trajectories and lengths $S$, or even varying operational modes of relative motion (rotating, reversing, reciprocating), a certain number of repetitive sliding cycles, etc. Many of these complex conditions can be set-up, controlled, and tested with so-called “tribometers” [Czichos, 2015].

Several authors already reported an improved tribological performance of LIPSS-covered surfaces on different materials, in various non-lubricated or lubricated tribological test conditions [Yu, 1999 / Mizuno, 2006 / Eichstädt, 2011 / Pfeiffer, 2013 / Bonse, 2014 / Gnilitskyi, 2019 / Ayerdi, 2019] and upon exploring some specific applications, including the improvement of mechanical seals [Chen, 2012], the optimization of the performance of cutting tools [Kawasegi, 2009], medical implant materials [van der Poel, 2019 / Kunz, 2020], etc. Summarizing this research, the basic ideas of using LIPSS for tribological applications are related to (i) a change in the geometrical contact area or roughness, (ii) a confinement of the lubricant in the valleys of the LIPSS being in tribological contact - acting as reservoir, (iii) an enhanced wetting of the lubricants at the laser-processed surfaces, or even laser-induced (iv) chemical or (v) structural surface modifications, such as oxidation or graphitization, that may (vi) additionally improve the efficiency of wear-reducing additives contained in many commercial lubricants. For more details on the tribological properties of LIPSS, the reader is referred to the pertinent literature here, such as Refs. [Bonse, 2018 / Bonse, 2021].

In the following, three specific examples of beneficial tribological performances of LIPSS-processed surfaces through a reduction of the CoF are presented in Figs. 24 to 27. These results demonstrate some prominent tribological test geometries/conditions, i.e., linearly reversing *ball-on-disk* tests (referred to as reciprocating sliding tribological tests, RSTT, see Fig. 24, right) and rotating *pin-on-disk* (Fig. 26, left) or *ring-on-disk* (Fig. 27(b)) tests.

Figure 24 summarizes the galvanometer laser-scanner-based processing of LIPSS on a $Al_2O_3$-$ZrO_2$-Nb dielectric-metal composite material (left). The fs-laser irradiation conditions were chosen to selectively form LSFL only on the elongated Nb-flakes of the composite (middle). These laser-processed samples were subsequently subjected to linear reciprocating sliding tribological tests (RSTT: 1 Hz reversing frequency, $F_N$ = 1.0 N normal force) against a 10-mm-diameter ball of polished $Al_2O_3$, employing a stroke length of $S$ = 1 mm, and a direction perpendicular to the elongated direction of the metal flakes in a Ringer solution as aqueous lubricant (right) [Kunz, 2020].

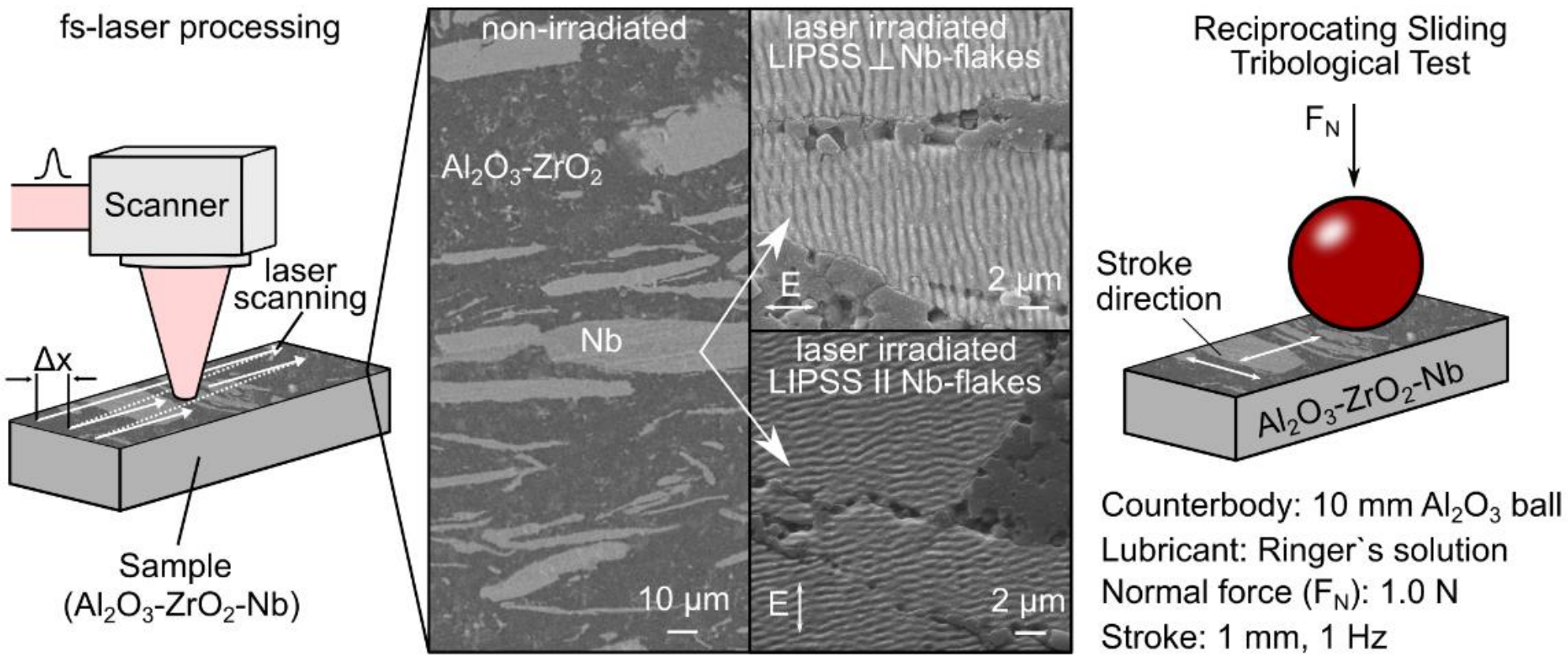

Fig. 24: Scheme of fs-laser processing and reciprocating sliding tribological tests (RSTT) on selectively LIPSS-structured $Al_2O_3$-$ZrO_2$-Nb composites. (Reprinted from [Kunz, 2020], Appl. Surf. Sci., Vol. **499**, Kunz et al., Tribological performance of metal-reinforced ceramic composites selectively structured with femtosecond laser-induced periodic surface structures, 143917, Copyright (2019), with permission from Elsevier).

The results of the RSTT for samples processed with a fs-laser [1025 nm, 300 fs, 1 kHz] at three different laser peak fluence levels along with that of a non-irradiated (polished) reference are presented in Fig. 25. The graphs show the CoF vs. the number of sliding cycles (here, up to 1000) for stroke directions either perpendicular to the LSFL ridges or parallel to it. In both cases, an up to a factor of three reduced CoF can be seen for the LSFL-covered Nb-flake samples when compared to the polished reference surface.

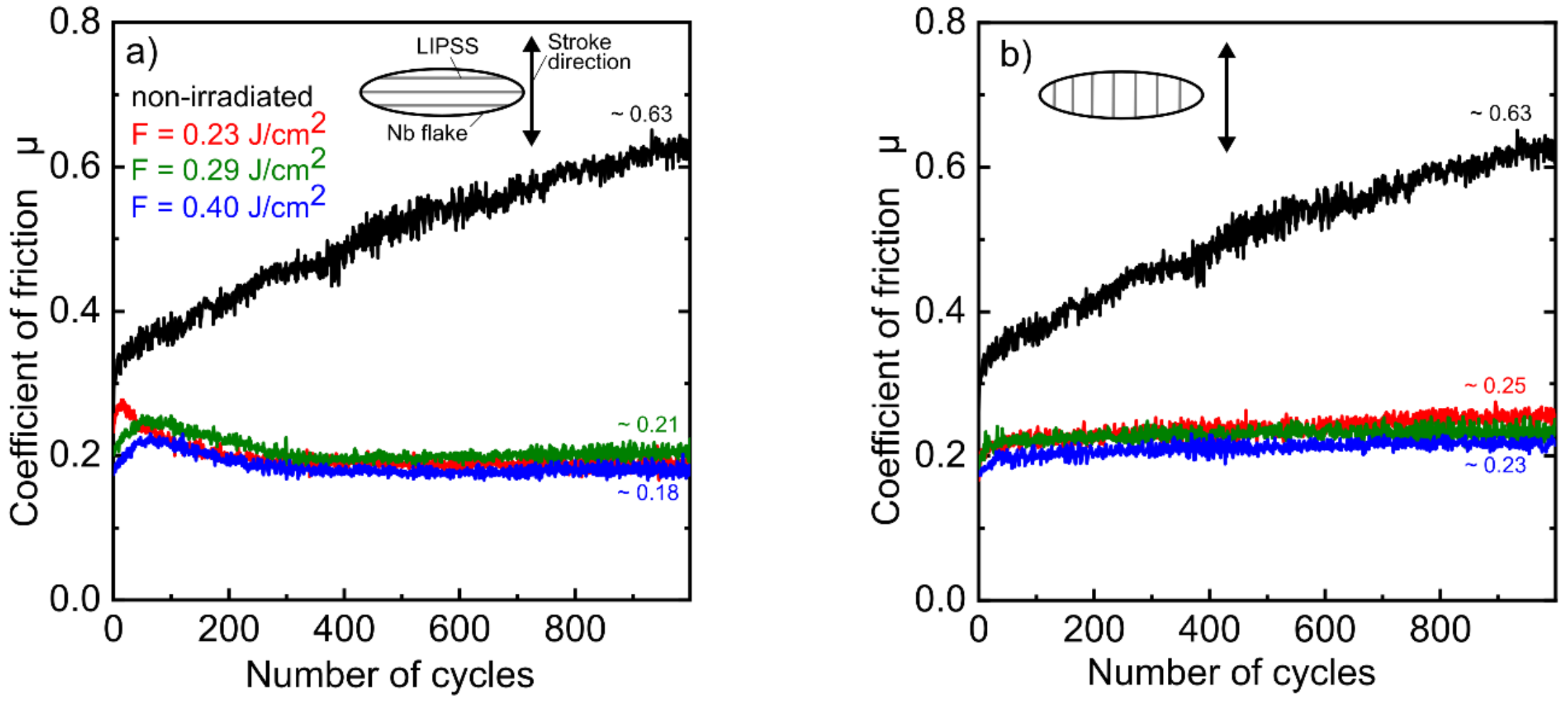

Fig. 25: Coefficient of friction (μ) vs. the number of RSTT sliding cycles for selectively LIPSS-structured $Al_2O_3$-$ZrO_2$-Nb composite surfaces (see Fig. 21) upon fs-laser processing with three different peak fluence values (0.23 J/cm$^2$, 0.29 J/cm$^2$, 0.40 J/cm$^2$) along with a reference of the non-irradiated material. The insets indicate the direction of the LIPSS (LSFL) with respect to the stroke direction used in the RSTT (a: perpendicular, b: parallel). (Reprinted from [Kunz, 2020], Appl. Surf. Sci., Vol. **499**, Kunz et al., Tribological performance of metal-reinforced ceramic composites selectively structured with femtosecond laser-induced periodic surface structures, 143917, Copyright (2019), with permission from Elsevier).

A similar beneficial tribological effect of LIPSS was reported by Gnilitskyi et al. [Gnilitskyi, 2019] for fs-laser processed [1030 nm, 213 fs, 600 kHz] initially polished stainless steel (X5CrNi1810) samples, when tested in a pin-on-disk test against a 2-mm-diameter pin of chrome steel (100Cr6) in a commercial engine oil as a lubricant [$F_N$ = 0.5 to 7.5 N, $v_{rel}$ = 7 mm/s]. Figure 26 shows the so-called *Stribeck curve* (CoF vs. Stribeck number) of the fs-laser processed and a polished surface, acquired in the regime of boundary lubrication to mixed lubrication. The Stribeck number $SN = \eta \times v_{rel}/P$ considers the dynamic viscosity $\eta$ of the oil, the relative velocity $v_{rel}$ between the pin and the rotating disc surface, and the areal pressure $P$. In the entire tested range of *SN*, a reduction of the CoF by 10% to 25% can be seen in Fig. 26(b) when comparing the results of the LSFL-covered surface (red open circles) with that of the polished reference (black open squares). Simultaneously, a reduction of the wear by ~65% was found [Gnilitskyi, 2019].

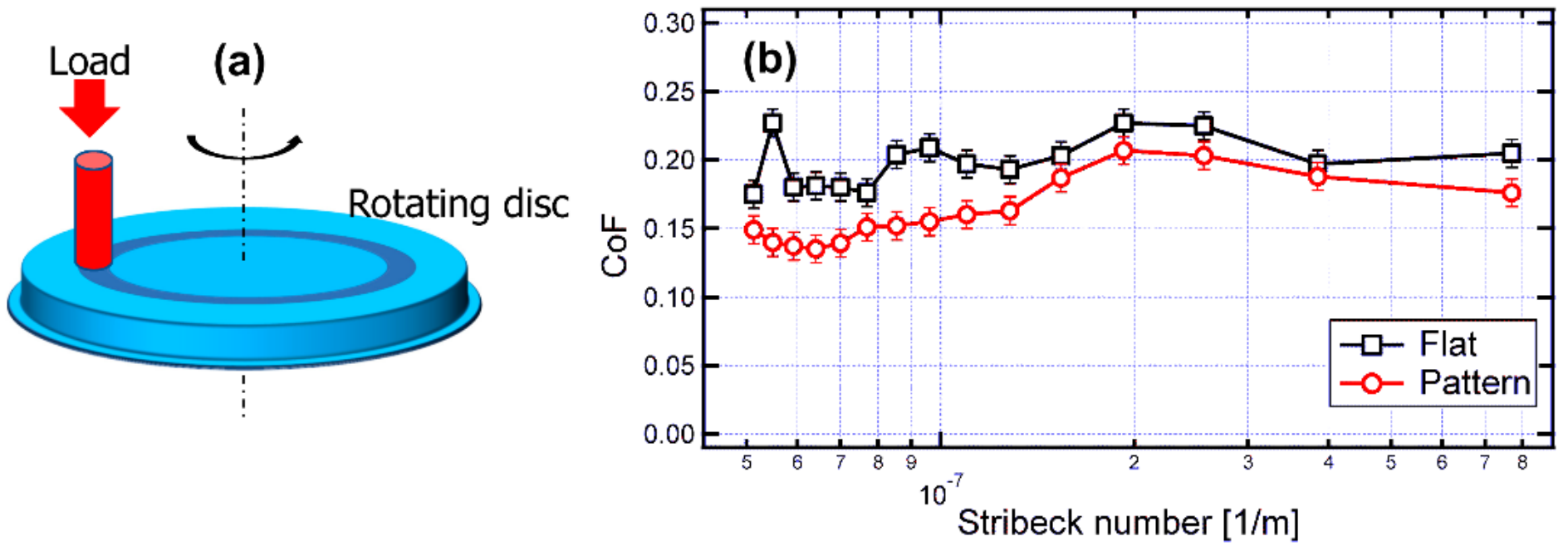

Fig. 26: (a) Scheme of pin-on-disk tribological test. (b) Stribeck curve of LIPSS (LSFL, referred to as "pattern") on stainless steel along with a reference curve for a polished (flat) surface. (Reprinted from [Gnilitskyi, 2019] Gnilitskyi et al., Tribological Properties of High-Speed Uniform Femtosecond Laser Patterning on Stainless Steel, Lubricants (Basel, Switzerland) **7**:83, Copyright 2019 under Creative Commons BY 4.0 license. Retrieved from https://doi.org/10.3390/lubricants7100083).

A beneficial effect through LIPSS was also reported by Chen et al. [Chen, 2012], studying the CoF of partially fs-laser-processed [800 nm, 120 fs, 1 kHz] initially polished silicon carbide (SiC) surface (Fig. 27(a)) in a ring-on-disk tribological test configuration in water environment (Fig. 27(b)). The CoF of such a fully textured SiC seal is ~20% smaller than that of a conventional polished one (Fig. 27(c)).

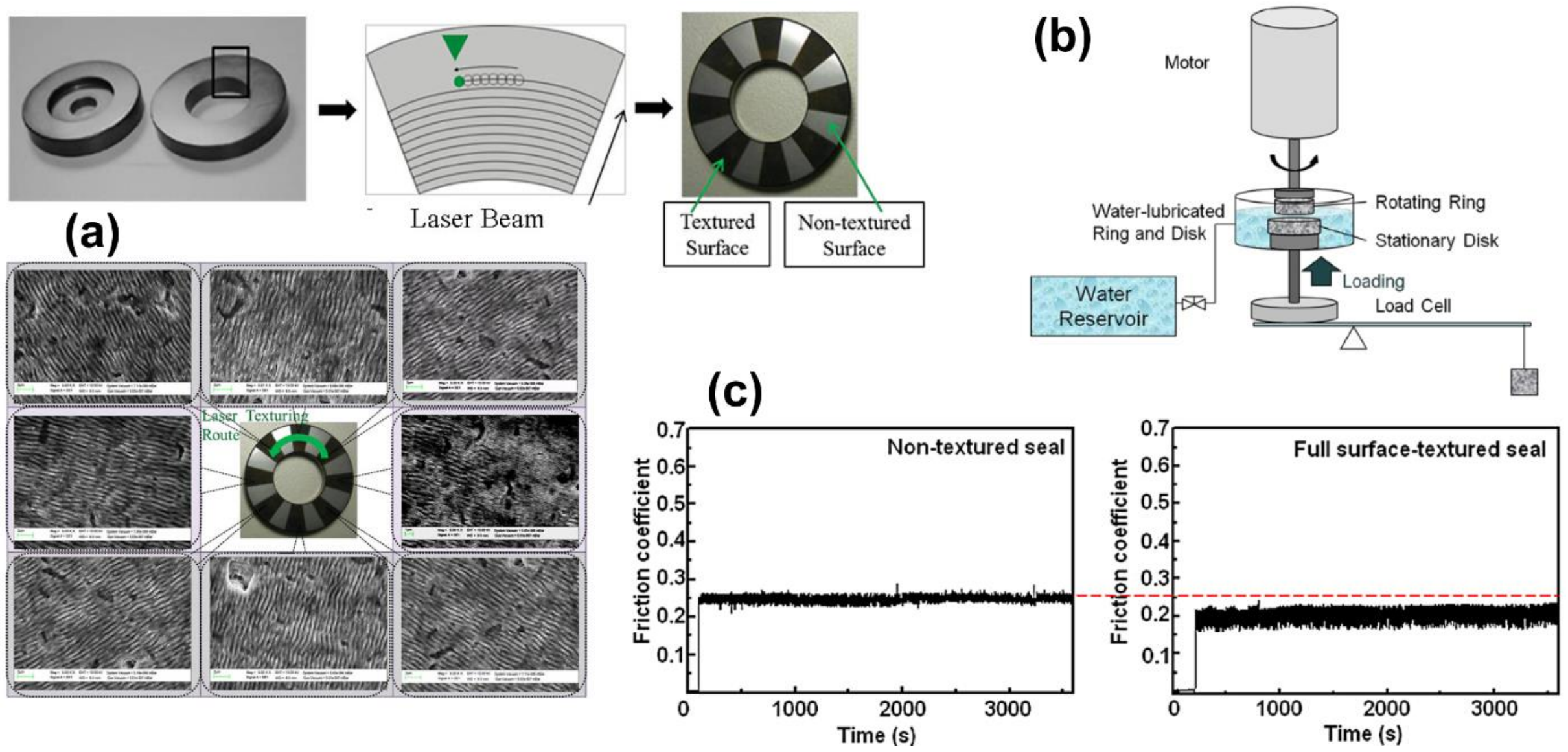

Fig. 27: (a) SiC seal consisting of two disks, of which one was fs-laser textured with LIPSS on eight individual segments. (b) Scheme of water-lubricated ring-on-disk tribometer. (c) CoF vs. operational time of the tribometer for a non-textured seal (left) and a LIPSS-textured seal (right). Image (a)-(c) taken from [Chen 2012]. (Reprinted by permission from Springer-Verlag: Applied Physics A **107**:345–350 (Microstructure and lubricating property of ultra-fast laser pulse textured silicon carbide seals, Chen, C.Y. et al.), Copyright (2012)).

## Management of Lubricants

Stark et al. [Stark, 2019] explored another tribological situation related to starvation lubrication conditions (Fig. 28). The authors demonstrated through in-situ optical microscopy and with a rotating ball-on-three-plates tribometer that parallel microchannels with periods of a few micrometers and depths below 1.5 µm generated by direct laser interference patterning (DLIP; 1030 nm, 6 ps, 400 kHz) on a stainless steel (X90CrMoV18) surface can induce a directed passive flow of lubricant (here polyalphaolefin oil, PAO, at room temperature) directly into the <250 µm diameter tribological contact area [Stark, 2019]. The effect is mainly driven by capillary forces of the lubricant within the microchannel surface structure (see also Section 2.2 *Liquid Management and Surface Wetting*).

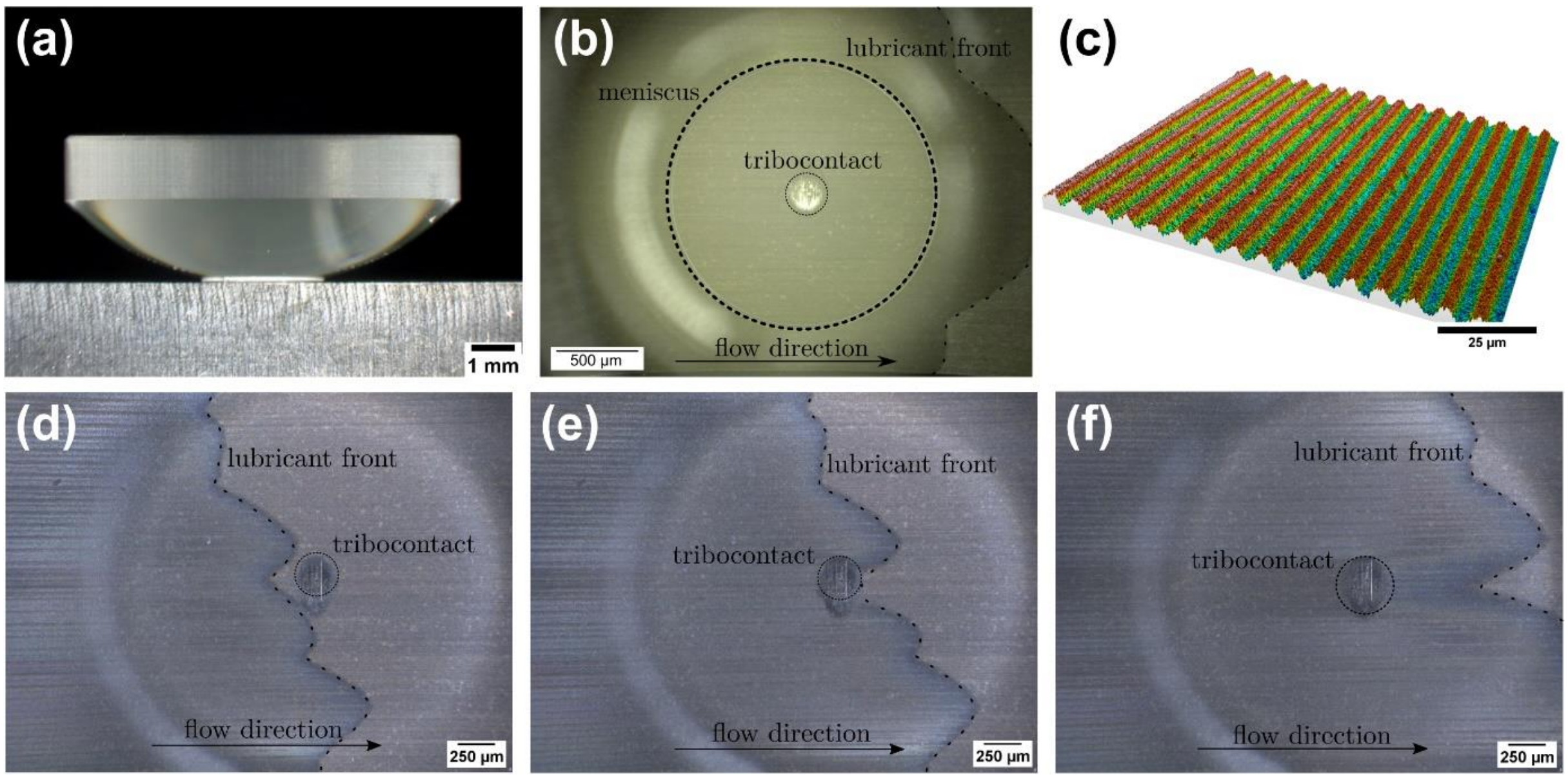

Fig. 28: (a) Side-view of the laser-processed sample (bottom) and a transparent glass lens (top) made for in-situ observation of the lubricant fluid flow into the tribo-contact (middle). (b) Through-the-lens microscopic imaging of the tribo-contact (central bright disk) centered in the meniscus of lubricant confined between the lens and the sample. (c) Surface topography of ps-DLIP generated parallel microchannels on stainless steel for a passive directed lubricant (PAO) transport. (d-f) Temporal evolution of the PAO transport into the tribo-contact area through capillary forces within the horizontally oriented microchannels. (Reprinted from [Stark, 2019], Stark et al., Avoiding Starvation in Tribocontact Through Active Lubricant Transport in Laser Textured Surfaces, Lubricants (Basel, Switzerland) **7**:54, Copyright 2019 under Creative Commons BY 4.0 license. Retrieved from https://doi.org/10.3390/lubricants7060054).

## 2.5 Energy Applications

The ever-growing world's energy needs constantly open multidisciplinary challenges for researchers and engineers when developing efficient methods and infrastructures to harvest, transport and storage energy. Laser micro- and nanoprocessing of materials via LIPSS constitute a versatile processing method with potential to surface functionalization for certain energy applications. The fabrication of LIPSS can modify surface properties by producing topographical changes (either to enhance optical absorption/transmission/scattering, to alter electronic properties such as the electric conductivity, or to increase the effective surface area for surface-based chemical processes) or by inducing localized chemical changes as a product of the modulated laser fluence imprint. In the following, a set of applications will be discussed where LIPSS influence the final material-device functionality including solar cells, heterogeneous conductive structures on indium-doped tin oxide (ITO) substrates, superconductive surfaces, and enhanced battery storage capabilities.

### Solar Cells

The fabrication of solar cells usually requires the assembling of at least three materials disposed as stacked layers: a top transparent electric contact, followed by a doped semiconductor, and finally a back electric contact. The processing of LIPSS improve the overall efficiency by modifying one of the layered materials in basically two ways: by modifying transparent surfaces to increase light transmission into the semiconductor active layer [Knüttel, 2013 / Liu, 2018b /

Nigo, 2020], or by inducing localized chemical changes in the semiconductor active material [Cui, 2016 / Rodriguez-Rodriguez, 2018].

One of the first experimental realizations of LIPSS on fully functional organic solar cells is presented in Fig. 29. The general thin-film solar cell architecture is illustrated in Fig. 29(a). In this case, laser irradiation from a pulsed ns-Nd:YAG laser with an output wavelength of 532 nm (second harmonic) is directed over Poly(3-hexylthiophene) thin films (P3HT) previously deposited onto an ITO substrate. For details on the specific layer chemical compositions and thicknesses, see [Cui, 2016]. The morphology of the LIPSS was characterized by atomic force microscopy (AFM) data, showing parallel polymeric line structures with heigh variations of about ~100 nm, and spatial periods corresponding to ~430 nm (Fig. 29(b)). The efficiency of regular and LIPSS-based solar cells was characterized by current vs. voltage (*I*-*V*) curves (Fig. 29(c)). From the reported data, such a LIPSS-corrugated solar cell features an overall increase of 10% efficiency with respect to the same solar cell composition/architecture without LIPSS.

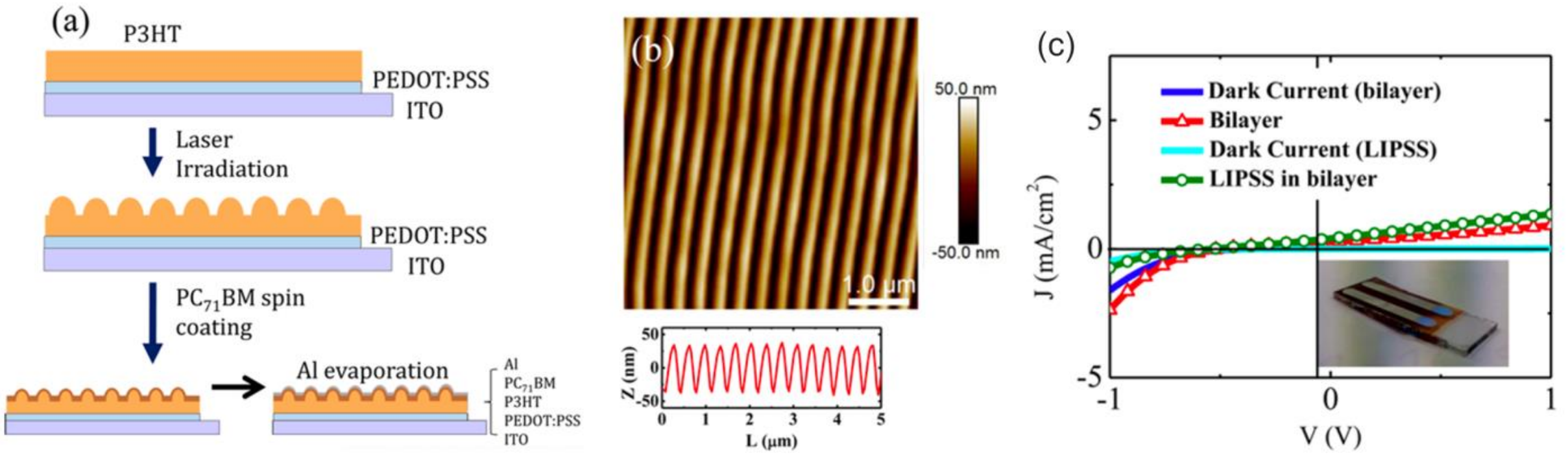

Fig. 29: (a) Scheme of the bilayer device architecture tested for photovoltaic characterization (b) Surface topography of the P3HT surface just after laser irradiation. A cross-sectional height profile of the LIPSS is provided at the bottom of the image. (c) *I*-*V* characteristics of the LIPSS solar cell in the dark and under illumination. Note that the current (*I*) is transformed here into the current density *J*. (Reprinted with permission from [Cui, 2016] (Cui et al., Laser-induced periodic surface structures on P3HT and on its photovoltaic blend with $PC_{71}BM$, ACS Appl. Mater. Interfaces **8**:31894–31901). Copyright (2016) American Chemical Society).

## Anisotropic Electric Conductivity

LIPSS fabrication on electrically conductive thin film surfaces has potential for the fabrication of individual parallel and periodic conductive surface-filaments that can be used for energy transport applications in the micro- to nanoscale. This potential has been reported on LIPSS fabricated on ITO thin films deposited on glass with electrical characterizations via AFM conductivity measurements [Gutierrez-Fernandez, 2020]. It was demonstrated that partial segregation of the conducting domains towards the valleys of the LIPSS could induce preferential (anisotropic) electrical conductivity along the LIPSS ridges.

The experimental realization with direct electrical characterization via macroscopic and microscopic 4-point probe resistivity measurements was reported in [Lopez-Santos, 2021]. Here, LIPSS with spatial periods of about ~900±50 nm are responsible for resistivity differences longitudinally ("Micro-L") or transversally ("Micro-T") to the LIPSS orientation. The anisotropy is naturally imposed by the LIPSS orientation, and a wide range of resistivities are feasible mainly due to changes in the crystalline structure of the ITO layers as a result of the femtosecond pulses interaction. The actual micro-characterization via the 4-point probe method is displayed in Fig. 30(a), where the microprobes are placed longitudinally and transversally to the LIPSS orientation. The current vs. voltage (*I*-*V*) curves for two areas fabricated by a low (LF: 0.64 J/cm$^2$) and a high (HF: 0.89 J/cm$^2$) peak laser fluence level are presented in Figs. 30(b) and 30(c), respectively. In both graphs, the slopes of the lines correspond to the electric resistivity in each case, showing clear differences between Micro-L and Micro-T characterization. In particular, for the low fluence sample, there is a slope difference of a factor of two, higher for the Micro-L configuration than for the Micro-T one. These finding suggest that the presence of anisotropic conductivity is induced due to a combination between the LIPSS topography and local resistivity changes due to the laser irradiation.

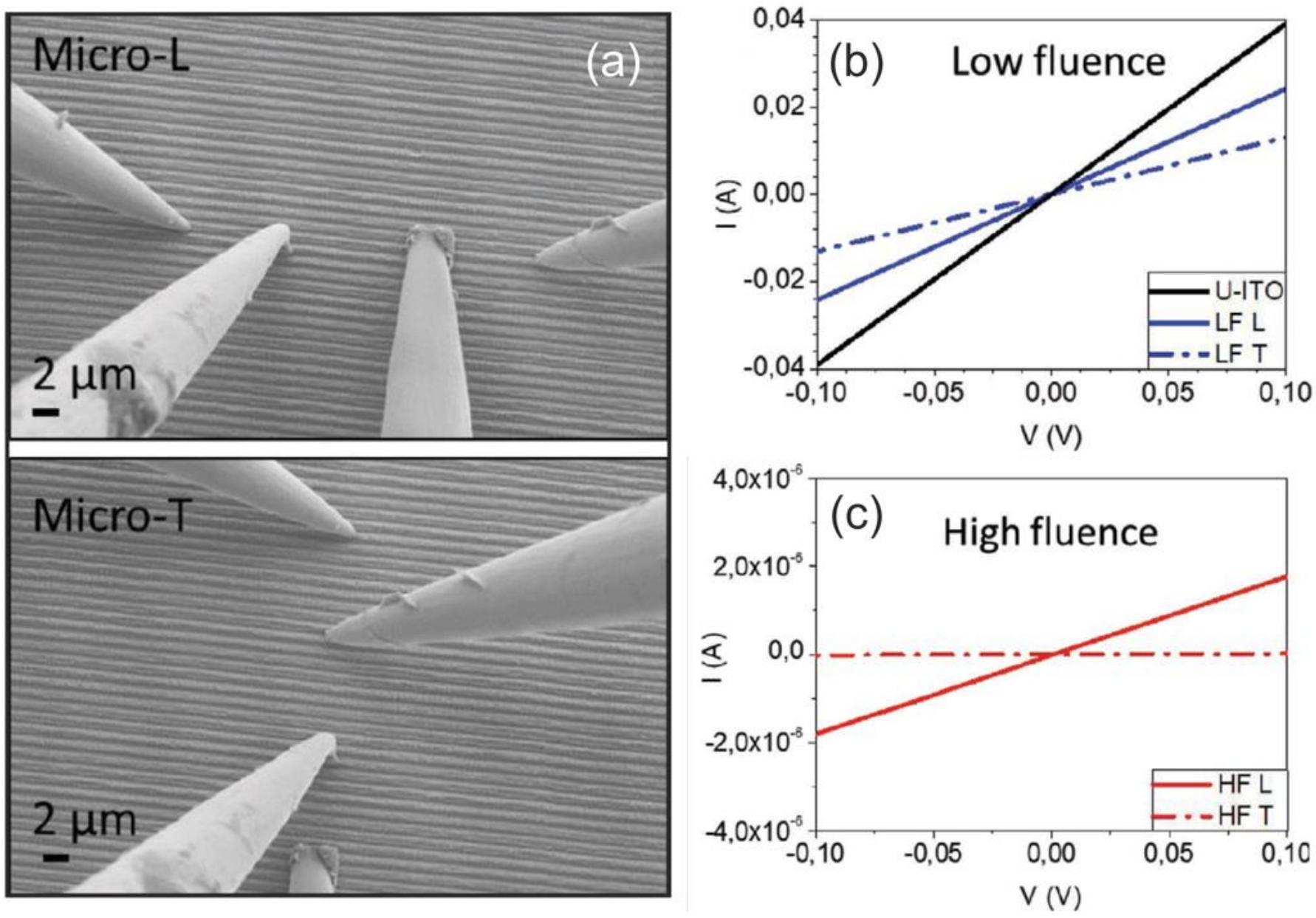

Fig. 30: (a) Micro-configuration 4-point probe measurements. The actual distance between the electrodes is determined from SEM images. Finger-tip-like metallic electrodes are carefully located at the ridges of the structures. (b), (c) *I-V* curves measured for the low fluence (LF) conditions in the macro-configuration and for high fluence (HF) conditions in the micro-configuration, with the electrodes aligned longitudinally (Micro-L) and transversally (Micro-T) to the LIPSS direction. The *I-V* curve of a non-irradiated film is included as reference in (b). (Reprinted from [Lopez-Santos, 2021], Lopez-Santos et al. Anisotropic resistivity surfaces produced in ITO films by laser-induced nanoscale self-organization, Adv. Opt. Mater. **9**:2001086, Copyright 2020 under Creative Commons BY 4.0 license. Retrieved from https://doi.org/10.1002/adom.202001086).

## Surface Superconductivity

Femtosecond laser irradiation for the fabrication of LIPSS can be used to modify the magnetic responses of a material. This opens widely unexplored possibilities in terms of the final surface properties that can be accessible through surface processing techniques based on LIPSS. Recently, it has been demonstrated that magnetic responses obey topographical and chemical changes induced by the laser treatments including LIPSS [Czajkowski, 2017 / Sanchez, 2020]. For applications where the material's magnetic response is a defining factor, such as surface superconductivity, the realization of LIPSS constitutes an important parameter to study.

The realization of LIPSS on superconductor materials such as niobium, has been reported in [Cubero, 2020a / Cubero, 2020b], where direct current (DC) magnetization and alternating current (AC) susceptibility measurements at cryogenic temperatures with external magnetic fields of different amplitude and orientation were analyzed and reported for samples with and without LIPSS, produced on different atmospheres, as indicated in Figs. 31(a)-31(d). The electric characterizations demonstrated a correlation of the surface superconducting properties and the LIPSS orientation. In fact, the upper critical field, $H_{c2}$, was found not to be affected by the LIPSS, but the surface critical field, $H_{c3}$, presented higher values when the external magnetic field ($H$) was set parallel to the LIPSS orientation (Fig. 31(e)). Hence, it is possible to define a preferred superconductivity direction by the orientation of the imprinted structures at the surface. Interestingly, despite the induced effects of laser irradiation on the surface superconducting properties, the $H_{c3}$ values are almost unaffected by differences in the LIPSS period [Cubero, 2020a].

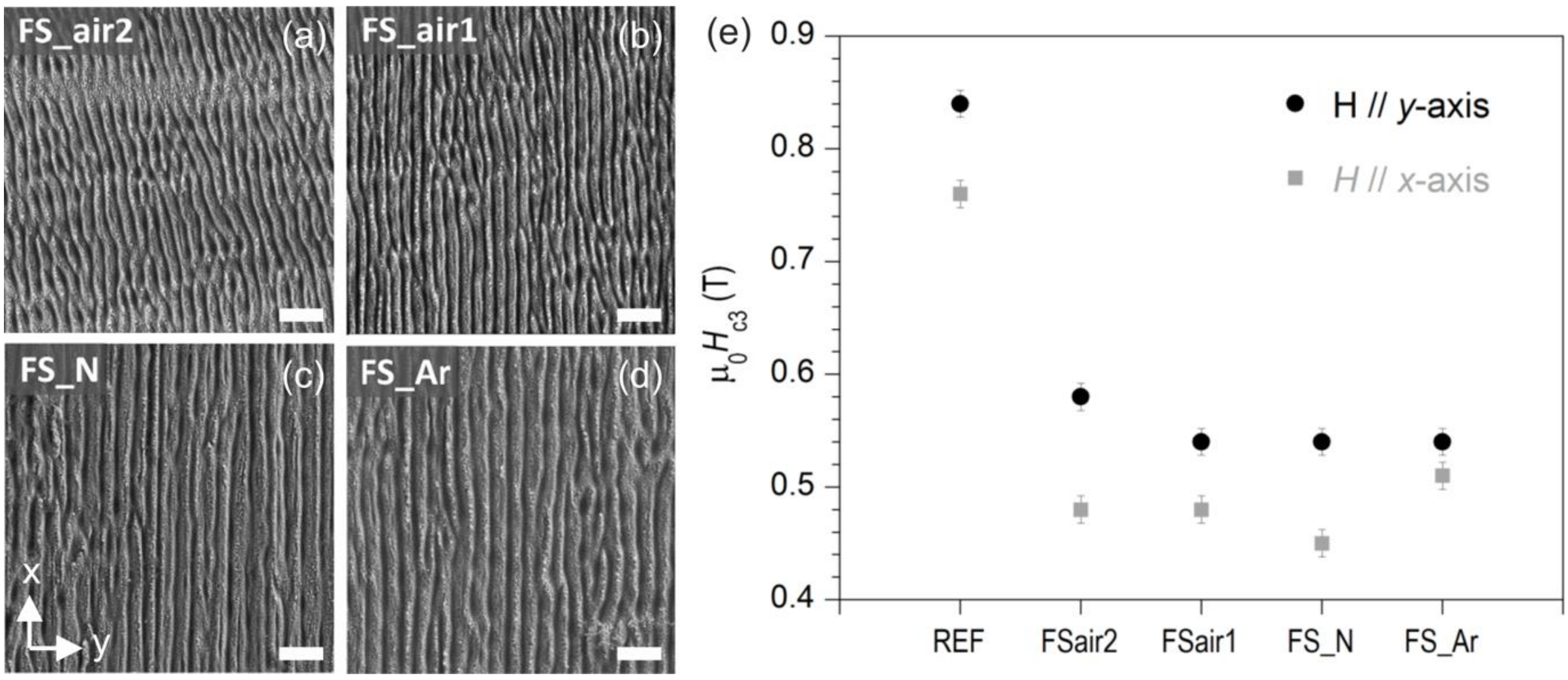

Fig. 31: SEM micrographs (secondary electrons) of the fs-laser-irradiated surfaces produced under different gaseous atmospheric conditions: (a) and (b) in air, (c) in nitrogen and (d) in argon. The scale bar corresponds to 2 µm for all micrographs. The rolling direction of the pristine Nb sheet material is parallel to the *y*-axis in all samples, except for FS_air2 in (a), where it is parallel to *x*. (e) Critical surface fields, $\mu_0 H_{c3}$, of all analyzed samples. (Reprinted from [Cubero, 2020a], Cubero et al. Surface superconductivity changes of niobium sheets by femtosecond laser-induced periodic nanostructures, Nanomaterials (Basel, Switzerland) **10**:2025, Copyright 2020 under Creative Commons BY 4.0 license. Retrieved from https://doi.org/10.3390/nano10122525).

## Energy Storage

The functionalization of nickel metal electrodes acting as a cathode material optimized via LIPSS for micro-batteries applications has been reported in [Neale, 2014]. It was demonstrated that the fabrication of LIPSS at throughputs of 4 $cm^2$/s constitutes a valuable and fast one-step approach for reliable and reproducible micro-scale surface modifications on nickel-based batteries electrodes.

The morphology of the fabricated LIPSS via SEM micrographs is displayed in Figs. 32(a)-32(d), showing structures with periods around ~1 µm. Cyclic voltammetry characterization with LIPSS-covered and non-irradiated nickel samples showed that the anodic peak current density of the laser processed sample at electric potentials between 0.35 V to 0.40 V (Fig. 32(e)) was over 2.5 times larger than the equivalent peak current density for the unprocessed nickel electrode (25.6 mA/$cm^2$ against 9.9 mA/$cm^2$). This is translated into net charge densities of 1.2 mC/$cm^2$ for the unprocessed nickel foil and 4 mC/$cm^2$ for the processed sample. This increase in electrochemical activity is attributed, on one hand, to an overall increase of the effective surface area relative to a non-modified nickel surface, and on the other, to chemical changes in the form of thicker oxide layers that may benefit the overall performance. Different fs- and ps-laser-processing conditions to produce the LIPSS were also studied, including laser wavelengths (532 nm, 775 nm and 1064 nm), in order to identify the influence of the different topographies/periods on the maximum peak current density achieved. It was demonstrated that the percentage of surface coverage influences the electrode performance, as it can be observed in the data plotted in Fig. 32(f).

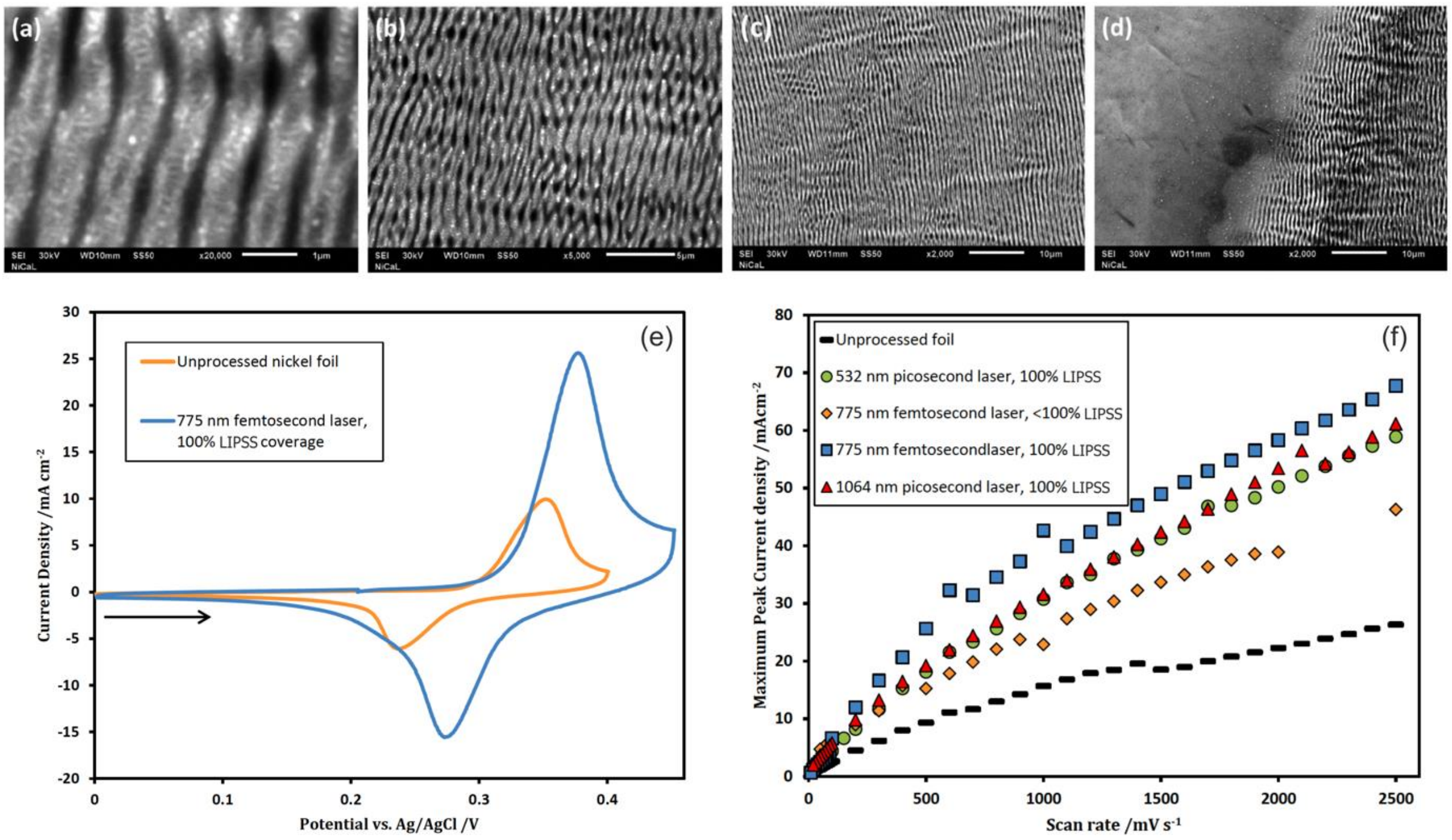

Fig. 32: (a)-(c) Top-view scanning electron micrographs of LIPSS fabricated with a laser wavelength of 1064 nm on nickel surfaces under different magnifications. (d) SEM micrograph centered in the boundary of laser-processing, showing a relatively planar non-irradiated nickel surface (left) and the LIPSS (right). (e) Cyclic voltammetry at 500 mV/s of a LIPSS-covered nickel electrode (blue curve) and a non-irradiated nickel foil electrode as reference (orange curve). (f) Anodic peak current density values for full range of scan rates and irradiation wavelengths. "% LIPSS" corresponds to the LIPSS-coverage across the nickel surface. (Reprinted from [Neale, 2014], Neale et al., Electrochemical performance of laser micro-structured nickel oxyhydroxide cathodes, J. Power Sources **271**:42, Copyright 2014 under Creative Commons BY 3.0 license. Retrieved from https://doi.org/10.1016/j.jpowsour.2014.07.167).

## 2.6 Other Technical Applications

This Section is dedicated to LIPSS applications, which do not fall in one of the other application areas mentioned above. It will be discussed, that LIPSS can improve the performance of piercing punch tools as well as the improvement of gasoline injection nozzles, and the catalytic activities of electrodes.

### Improved Piercing Punch Tool Performance

It was demonstrated by Aizawa et al. [Aizawa, 2020], that the LIPSS nanostructure pattern processed on the curved surface area of a diamond-coated cylindrical WC (Co) punch tool was transcribed onto the side wall of the concurrent metal sheet piercing hole during the punching process, which finally lead to a smoother metal sheet hole surface. Figure 33 illustrates SEM micrographs of the diamond-coated punch tool without (a) and with LIPSS (b), as well as images of the pierced metal hole surface after punching it without (c) and with LIPSS (d).

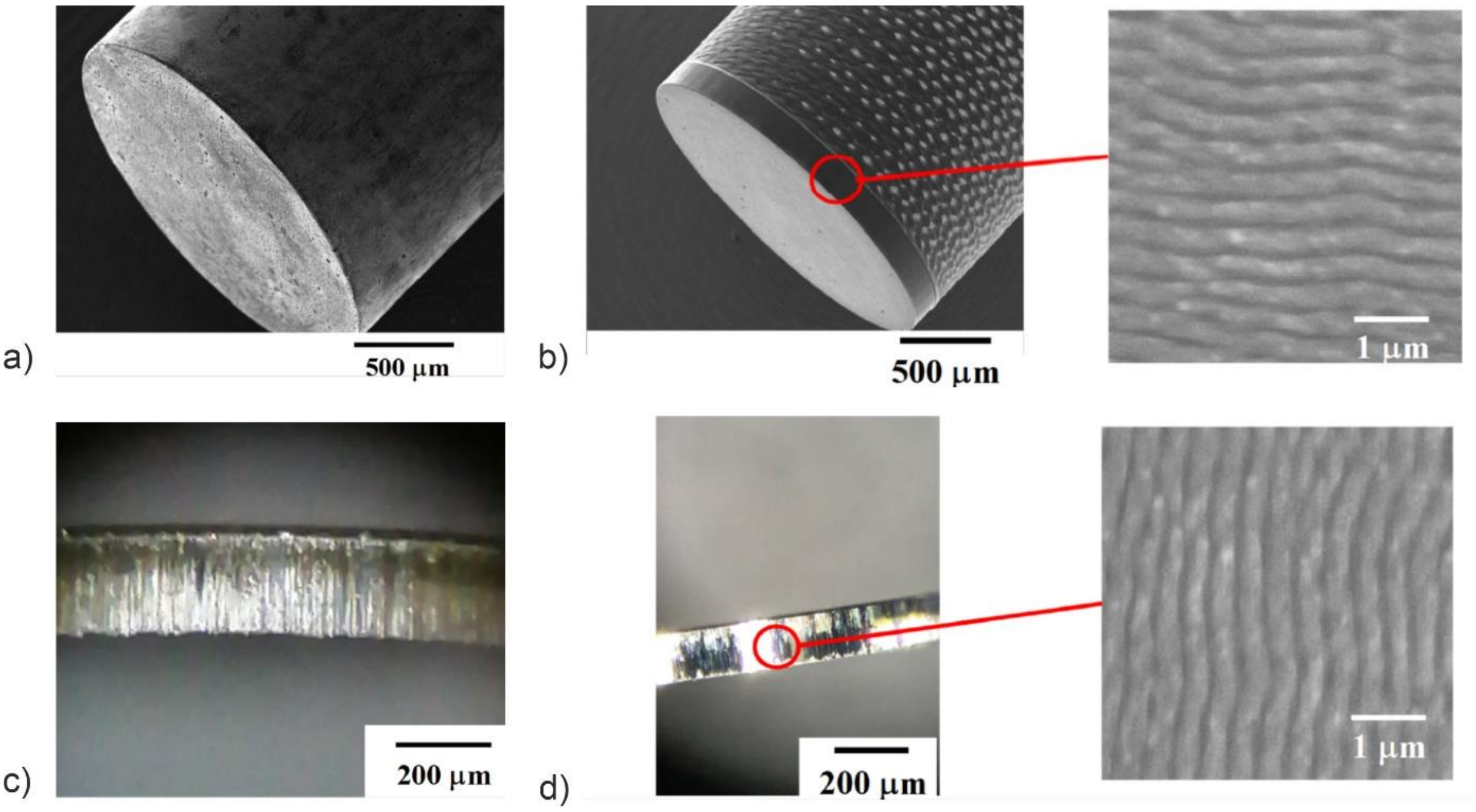

Fig. 33: (a) SEM micrograph of unstructured CVD (Chemical Vapor Deposition)-diamond coated WC punch tool; (b) SEM micrograph of laser-structured CVD (Chemical Vapor Deposition)-diamond coated WC punch tool with magnified inlet of LIPSS; (c) side-view of hole with rough surface punched by unstructured WC punch tool; (d) side-view of hole punched by laser-structured WC punch tool with smooth area with magnified inlet of SEM micrograph showing imprinted microstructures onto the punched side wall. (Reprinted from [Aizawa, 2020], Aizawa et al., Simultaneous Nano-Texturing onto a CVD-Diamond Coated Piercing Punch with Femtosecond Laser Trimming, Appl. Sci. (Basel, Switzerland) **10**:2674, Copyright 2020 under Creative Commons BY 4.0 license. Retrieved from https://doi.org/10.3390/app10082674).

## Improved Gasoline Injection Nozzles

It was demonstrated, that LIPSS in the spray hole of gasoline direct injection nozzles enhance the atomization process of the fuel by decreasing the average droplet size of about 10% and by simultaneously increasing the spray angle by 20% with respect to regular laser trepanned holes [Romoli, 2014]. Figure 34 shows SEM micrographs of a cross-section of a fs-laser drilled spray hole with the corresponding LIPSS on the inner surface of the hole [Romoli, 2014]. Furthermore, fs-laser drilled spray holes covered with LIPSS consist of smaller surface features than compared to micro-Electrical Discharge Machining (μ-EDM) or Laser Micro-Jet (LMJ) machining, at which the process related craters (μ-EDM) and grooves (LMJ) are in the scale of several micrometers, and produces the sharpest edges with a radius of less than 1 μm, while the μ-EDM edge radius typically lies between 2 and 4 μm and LMJ produces unwanted and uncontrollable protrusions [Romoli, 2015].

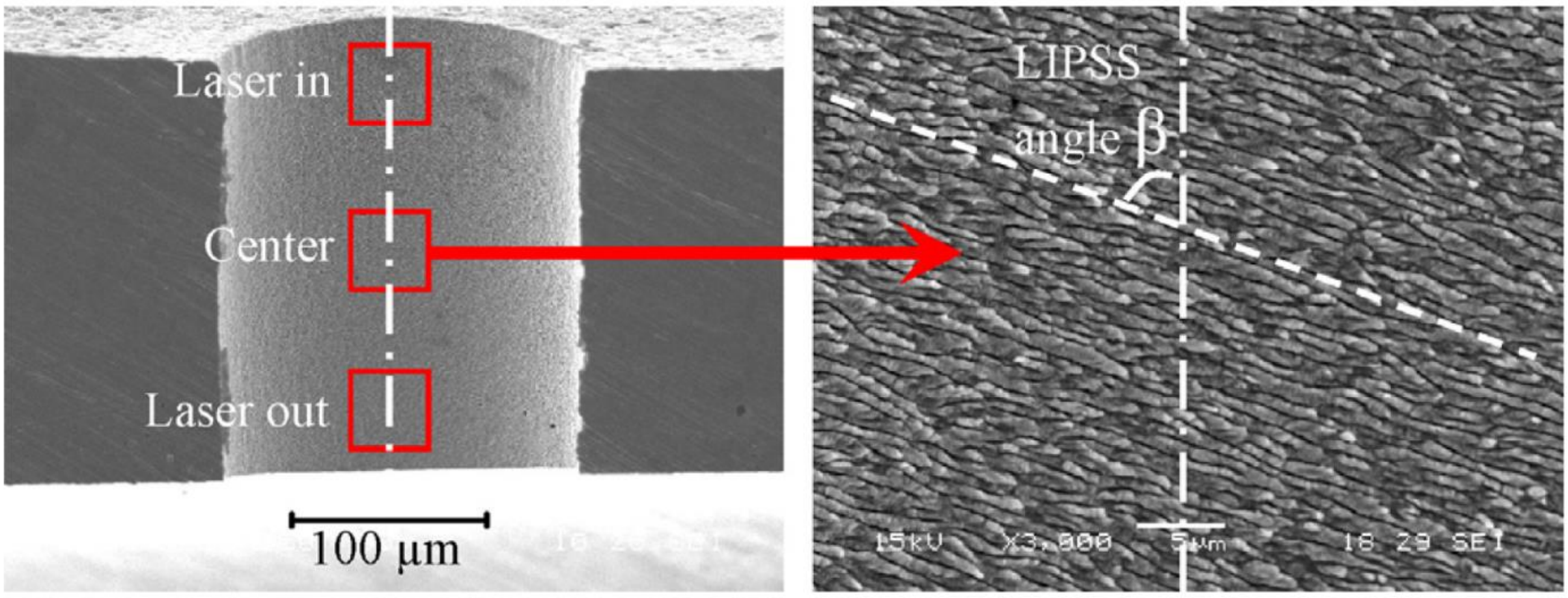

Fig. 34: SEM micrographs of a cross-section of a fs-laser drilled spray hole (left) and the magnified nanostructures in the center of the surface of the spray hole (right). (Reprinted from [Romoli, 2014], CIRP Annals, Vol. **63**, Romoli, et al., Ultrashort pulsed laser drilling and surface structuring of microholes in stainless steels, 229–232, Copyright (2014), with permission from CIRP).

### Increased catalytic activity

The fabrication of LIPSS typically leads to an increase of the nanoscale surface roughness and the total surface area. Therefore, LIPSS can enhance the catalytic activity [Neale, 2014 / Lange, 2017], depending on the specific interfacial interactions and chemical reactions. It was found by Lange et al. that the electrochemical activity in the oxygen reduction reaction on laser processed platinum plates increased by a factor of 1500 [Lange, 2017].

# 3. Industrialization

This sub-Chapter sheds some light on the current state of industrialization of LIPSS by briefly analyzing the available market segment (Section 3.1 *Market Analysis*), providing examples for estimating the production costs of LIPSS based on the currently available technology (Section 3.2 *Production Costs*), and giving a survey on patent applications that are relying on LIPSS (Section 3.3 *Patent Situation*).

## 3.1 Market Analysis

Although LIPSS and their applications have been extensively studied, their practical manufacturing has not yet emerged from an experimental status (*Technology Readiness Levels* of TRL 3 to 5) into industrial applications (TRL > 5). A main reason for this obstacle is the relatively low areal processing rate when scanning the pulsed laser beam across the surface of the workpiece via galvanometer scanners, which typically offer scan velocities up to 10 m/s in a line. However, upcoming modern laser systems with high average laser powers in the kilowatt regime with pulse repetition frequencies in the megahertz range in combination with polygon scanners with scan velocities up to 1000 m/s [Streek, 2019] open the door for achieving industrial rate processing times [Fraggelakis, 2017 / Schille, 2020]. Currently, the areal LIPSS processing rates are at the $m^2$/min level [Schille, 2020]. It can, however, be expected that future improvements in ultrafast laser technology that are enabling high energy laser pulses at (burst) repetition frequencies in the Gigahertz range [Mishchik, 2019] along with advanced laser scanner technologies and in combination with smart scanning strategies for controlling the residual heat load to the laser irradiated material, will enable areal LIPSS processing rates at the $m^2$/s level [Bonse, 2020b].

Moreover, it must be noted that the specific LIPSS processing parameters presented in most scientific publications were not optimized for enabling the lowest overall processing time for the used laser wavelength and the given material. A method to predict the laser processing parameters for optimized production rates given the used laser and material was provided by Mezera et al. [Mezera, 2019].

Nevertheless, there are many applications that do not rely on large area processing but on the tailored marking of goods, tools, etc. One example is given by medical tools used in surgery that must be trackable during their entire lifetime, e.g. by marking serial numbers, bar codes, or data matrix codes with high optical contrast. For that application, an industrial process relying on ultrashort laser-generated LIPSS named "black marking" was already established by the industry. The ultrafast laser blackened surface regions feature good corrosion resistance when a chromium containing steel is used. [Neugebauer, 2017].

In regard of the rapid developments on the laser market it is worth noting that the average power of ultrashort lasers is obeying an analogon to the famous Moore's law that was framing the developments in the field of computer technology for decades: a doubling of complexity every two years. More precisely, [Han, 2021] underlined that since the year 2000, the increase of the average laser power of ultrashort (sub-ps) pulsed lasers can be described by a power law, i.e., a growth rate $\sim 2^{Q/2}$ with $Q$ being the number of the years from beginning of the trend – consistently with the data presented in [Schille, 2021].

It is difficult to establish accurate forecasts toward a precise market analysis. However, the prices for ultrafast laser have been dropping in recent years [StrategiesUnlimited, 2020] and, just as a landmark, the world-wide annual growth rate of the revenue created by lasers in the market segment of fine metal processing is estimated to be 9.4% during the next 5 years, raising from a level of ~420 million USD in 2019 up to ~700 million USD in 2025 [StrategiesUnlimited, 2020]. Considering the wide fields of applications, the emerging up-scaling of the LIPSS manufacturing process to industrial demands, the dropping costs for ultrashort pulsed lasers and scanners and that first companies offering laser-based surface functionalization as a commercial service, it is only a matter of time until LIPSS are a standard industrial manufacturing technique.

## 3.2 Production Costs

In order to estimate the overall fabrication costs per area required to fabricate LIPSS, it is necessary to consider the individual costs for a basic set of devices and optical elements needed for the configuration of a standard laser direct laser writing setup. Usually, an ultrashort pulsed laser source provides a continuous train of pulses that will be directed to the surface of the material of interest after passing through a focusing element. Ultimately, the deposited energy dose (sometimes referred to as *accumulated laser fluence*) is given by product of the laser fluence $\phi_0$ and the effective number of laser pulses $N_{eff}$ that are hitting the surface per spot area unit ($\phi_{acc} = \phi_0 \times N_{eff}$). For a single scanned line, the latter is given by a relation between the laser pulse repetition frequency and the scanning speed for a given laser spot geometry at the focus (see Eq. (1)). In traditional systems, the laser beam remains static and translation stages position the sample at user defined positions and speeds. However, the increasing availability of high-repetition-rate ultrashort lasers impose a minimum laser scanning speed that is hard to reach by regular mechanical translation stages. For this reason, high-repetition-rate lasers are usually combined with fast laser scanners based on a set of individually controlled galvanometric or on polygonal mirrors. The laser scanning systems are subsequently coupled with F-theta lenses to guarantee the same focusing conditions even over extended surface areas.

Table 1 summarizes the general requirements and estimated prices for the given elements. It is important to note that the operational costs (physical space required, energy consumption, insurances, local regulations taxes, and operator salary) are not included in this estimation, mainly because these values depend on the country where the laser processing system is located. A generalized total cost estimation over a return-on-investment period of 5 years for 3-axes laser processing systems based on galvanometric or polygonal mirrors currently is ~254,000,- € and ~294,000,- €, respectively.

| Category | Element/Device | Features | Average prize estimation (2021) |
|---|---|---|---|
| **Laser system (Ls)** | High-repetition rate ultrashort laser | IR wavelength, Rep. frequency < 5 MHz, Pulse Energy < 100 µJ, Pulse duration ~500 fs | ~180,000 € |
| **Scanner (Sc)** | a. Galvanometric mirror-based scanner + F-theta lens with AR coating | <15 m/s | ~10,000 € |
| | b. Polygonal mirror-based scanner | 48 – 256 m/s | ~50,000 € |
| **Safety (Sf)** | Radiation safety and dosage control | Lead blinds, passive dosimeters | ~5,000 € |
| | Air filtering, exhaust system | Mobile station | ~3,000 € |
| | Eye protection | Safety goggles | ~1,000 € |
| **Maintenance (Ma)** | General maintenance and technical service | Filters, chiller refrigeration liquids, eventual system optimization | ~5,000 €/year |
| **Miscellaneous (Ms)** | Optical elements | Mirrors, mounts | ~5,000 € |
| | Translation stages | Independent *X-Y-Z* axes | ~5,000 €/per axis |
| | Optical table | Optical table with active isolation legs | ~10,000 € |

Tab. 1: Elements needed for a general laser processing system and its estimation price on the market (Information consulted on May 26th, 2021).

The relation between the number of effective pulses per beam spot diameter ($N_{\mathrm{eff_1D}}$), the laser beam waist radius at the material surface ($w_0$), the laser pulse repetition frequency ($\nu$) and scan velocity ($v_{\mathrm{scan}}$) is generally described by the Eq. (1):

$$N_{\mathrm{eff_1D}} = \frac{2 * w_0 * \nu}{v\mathrm{scan}}, \qquad (1)$$

According to [Florian, 2018], it is possible to fabricate low spatial frequency LIPSS (LSFL) on steel with a high-repetition rate laser using $N_{\mathrm{eff_1D}} = 10$. With this value and considering a laser beam waist radius $w_0 = 50$ µm, and a galvanometric scanner operating at $v_{\mathrm{scan}} = 15$ m/s, the maximum laser repetition frequency is $\nu = 1.5$ MHz according to Eq. (1). Under these conditions, it is theoretically possible to reach throughputs (areal processing rates) of 0.45 $m^2/h$. On the other hand, regular polygonal scanners operating at 50 m/s with the same beam waist at the surface allow for a laser repetition frequency of $\nu = 5$ MHz while keeping $N_{\mathrm{eff_1D}} = 10$, resulting in throughputs of 1.5 $m^2/h$. Note that for polygonal scanners operating at 950 m/s, recently processing rates of ~1.5 $m^2$/min were demonstrated for large-area surface texturing of metals [Schille, 2020].

For this calculation we assumed irradiations over a squared area of 1 m × 1 m accounting a total of 20,000 parallel lines of 1 m in length and 100 µm in width, all separated at a line distance $\Delta = 50$ µm (50% lateral beam overlap). Also, for the mirrors positioning times, 100 ms for polygonal mirrors and 333 ms for galvanometric mirrors (~5 times the processing time, 1 m/$v_{\mathrm{scan}}$) per line were assumed.

For an estimation of the total laser-processing time, we consider a time $t_{\mathrm{year}}$ corresponding to the average number of working days per year, e.g. $t_{\mathrm{year}} = 260$, and we assume that the system operational (production) time is in average $t_{op} = 18$ h per day. In order to estimate the *"Photonic costs"* to produce LSFL under the above specified conditions following a similar calculation as in [Bonse, 2015], a return-on-investment period of 5 years ($p = 5$) was considered. The formula for this calculation is included in Eq. (2) and considers the cost listed in Table 1:

$$\text{Photonic costs } [€/\mathrm{m}^2] = \frac{\text{Total setup price}}{\text{Total throughput}} = \frac{Ls + Sc + Sf + Ma * p + Ms}{\left(\frac{1200 * w_0 * \nu * \Delta}{N_{\mathrm{eff_1D}}} \frac{\mathrm{s}}{\mathrm{h}}\right) * t_{op} * t_{year} * p}, \qquad (2)$$

In conclusion, for a typical current ultrashort laser processing system that uses a galvanometric scanner, the photonic costs for the processing of LSFL are ~24 €/$m^2$, and for a polygonal scanner ~9 €/$m^2$.

## 3.3 Patent Situation

The technological advancements for industrial interests are usually considered for publication as a patent rather than scientific papers, in order to grant intellectual property rights to the inventors and developers of such technology for protecting their business. Importantly, patent publications are distinctly different from scientific literature, especially because of the economic and legal interests that are at play. Since the granting of a patent is country specific, and there are different stages that such patent documents will pass, we focus our attention here in publicly accessible patent applications - regardless of its legal approval status. The following Table 2 lists a number of published patents where LIPSS are used. Wherever possible, just the international WO patent number is provided. If not available, the European or national patent numbers are given. Regarding the use of the LIPSS, the type of application is additionally specified.

| Patent number | Title | Application | Publication date |
|---|---|---|---|
| WO2004035255 | Cyclic structure formation method and surface treatment method | LIPSS fabrication | 29/4/2004 |
| WO2006027850 | Method for enhancing adhesion of thin film | Surface adhesion | 16/3/2006 |
| WO2006065356 | Applications of LIPSS in polymer medical devices | Antibacterial surfaces | 22/6/2006 |
| WO2007012215 | Method and device for the defined structuring of a surface with a laser unit | LIPSS fabrication | 1/2/2007 |
| WO2008097374 | Ultra-short duration laser methods for the nanostructuring of materials | LIPSS fabrication | 14/8/2008 |
| CN101319347 | Method for crystal surface self-organizing growth of fine-nano-structure with femtosecond laser | LIPSS fabrication | 10/12/2008 |
| US20080299408 | Femtosecond laser pulse surface structuring methods and materials resulting therefrom | LIPSS fabrication | 04/12/2008 |
| WO2009090324 | Method and device for marking a surface using controlled periodic nanostructures | Marking | 23/7/2009 |
| WO2009117451 | Laser-based material processing methods and systems | LIPSS fabrication | 24/09/2009 |
| WO2012019741 | Method for producing embossing tools for microstructure elements using ultra-short laser pulses | Surface adhesion | 16/2/2012 |
| EP2692855 | Surface structuring for cytological and/or medical applications | Antibacterial surfaces | 17/7/2013 |
| US2014083984 | Formation of laser induced periodic surface structures (LIPSS) with picosecond pulses | LIPSS fabrication | 27/3/2014 |
| WO2014102008 | Method for producing structures on a surface of a workpiece | LIPSS fabrication | 3/7/2014 |
| WO2014111697 | Measurement scale | Optical | 24/7/2014 |
| WO2014111696 | A method of reading data represented by a periodic, polarising nanostructure | Data storage | 24/7/2014 |
| KR1427688* | Laser processing device and processing method | LIPSS fabrication | 1/8/2014 |
| RU2544892 | Method of producing micro- and nanostructures of surface of materials | LIPSS fabrication | 20/3/2015 |
| US20150136226 | Super-hydrophobic surfaces and methods for producing super-hydrophobic surfaces | Surface wetting | 21/5/2015 |
| WO2016039419 | Method for forming surface structure of zirconia-based ceramic, and zirconia-based ceramic | LIPSS fabrication | 17/3/2016 |
| WO2016090496 | Laser-induced metallic surface colouration processes, metallic nanoscale structures resulting therefrom and metallic products produced thereby | Optical | 16/06/2016 |
| WO2017053198 | Femtosecond laser-induced formation of single crystal patterned semiconductor surface | LIPSS fabrication | 30/3/2017 |
| WO2017063844 | Method for producing a scattering optical unit, scattering optical unit, and lamp having a scattering optical unit | Optical | 20/4/2017 |
| WO2017153750 | Method of reducing photoelectron yield and/or secondary electron yield of a ceramic surface; corresponding apparatus and product | Surface enhancement | 14/09/2017 |
| WO2018010707 | Method and system of ultrafast laser writing of highly-regular periodic structures | LIPSS fabrication | 18/1/2018 |
| US20180117797 | Methods of making hydrophobic contoured surfaces and hydrophobic contoured surfaces and devices made therefrom | Surface adhesion | 3/5/2018 |
| WO2018100952 | Method for manufacturing transfer mold roll having fine periodic structure and transfer mold roll | Surface adhesion | 7/6/2018 |
| US 10,876,193 | Method and apparatus for structural coloration of metallic surfaces | Optical | 1/1/2019 |
| CN109132998 | Method for inducing periodic surface structure of transparent dielectric material with single-pulse nanosecond laser | Wetting and cell adhesion | 4/1/2019 |
| US20190054571 | Nanosecond laser-based high-throughput surface nano-structuring (NHSN) process | LIPSS fabrication | 21/2/2019 |
| US20190168905 | Beverage container with internal antimicrobial texture | Antibacterial surfaces | 6/6/2019 |
| WO2019141626 | Mold for injection molding processes and molding process using the mold | Surface adhesion | 25/7/2019 |
| WO2019166836 | Using lasers to reduce reflection of transparent solids, coatings and devices employing transparent solids | Optical | 6/9/2019 |
| WO2020003090 | A component of a mold for molding preforms, a mold for molding preforms and a process for obtaining the component | Surface adhesion | 2/1/2020 |
| WO2020012137A1 | Process for nanostructuring the surface of a material by laser; assembly allowing this process to be implemented | LIPSS fabrication | 16/1/2020 |
| WO2020039217 | Laser fabricated superoleophilic metallic component with oil retention properties for friction reduction | Friction reduction | 27/2/2020 |
| KR1020200053368 | Metal surface color implementing method and metal plate with surface color implemented thereon | Optical | 18/5/2020 |
| CN111168233 | Method for inducing periodic structure on surface of optical glass by picosecond laser | Optical | 19/5/2020 |
| WO2020104370 | Sensor device and method for producing a sensor device | Surface wetting | 28/5/2020 |
| CN111250874 | Method of multi-pulse picosecond laser-induced semiconductor material periodic surface structure | LIPSS fabrication | 9/6/2020 |
| US20200198063 | Method for manufacturing substrate with transparent conductive film, substrate with transparent conductive film, and solar cell | Optical | 25/6/2020 |
| KR102131165 | Laser marking method using surface reflectance control | Optical | 7/7/2020 |
| US 10,786,874 B2 | Femtosecond laser pulse surface structuring methods and materials resulting therefrom | LIPSS fabrication | 29/09/2020 |
| US 10,876,193 B2 | Nanostructured materials, methods, and applications | LIPSS fabrication | 29/12/2020 |
| WO2021041166 | Elements for mitigating electron reflection and vacuum electronic devices incorporating elements for mitigating electron reflection | Optical | 4/3/2021 |
| WO2021048704 | Method of gluing metal parts | Surface adhesion | 18/3/2021 |
| US20210121983A1 | Process for nanostructuring the surface of a material by laser context and technological background | LIPSS fabrication | 25/3/2021 |

Tab. 2: Compilation of patents (Patent number, title, type of application, publication date) that include the formation of LIPSS within its description and the claims included in the patent application. The list is ordered by the publication dates ranging from 04/2004 to 04/2021.

# 4. Outlook

More than 50 years after their discovery, the topic of LIPSS moved from the field of fundamental research towards a pre-industrial stage that is currently screening systematically a plethora of possible applications - all are enabled through a remarkable variety of different surface functions of LIPSS (see Fig. 5 to recall the content of this Chapter). In view of the rapid ongoing progress of laser- and scanner-technology that further reduces the production costs of LIPSS and simultaneously boost up the manufacturing rates, it can be expected that during the coming five years, some industrial applications of LIPSS will be realized that will enter our daily life.

**Acknowledgements**

The authors acknowledge the projects Laser4Fun, CellFreeImplant, BioCombs4Nanofibers, and LaserImplant. These projects have received funding from the European Union's Horizon 2020 research and innovation programme under grant agreements No. 675063 (Laser4Fun, https://www.laser4fun.eu), No. 800832 (CellFreeImplant), No. 862016 (BioCombs4Nanofibers, http://biocombs4nanofibers.eu), and No. 951730 (LaserImplant, http://laserimplant.eu). C.F. acknowledges the Individual Fellowship Marie Skłodowska-Curie project FOCUSIS under grant agreement No. 844977.